\documentclass[prd,onecolumn,floatfix,nofootinbib]{revtex4}
\usepackage{xcolor}
\usepackage{subfigure}
\usepackage{float}
\usepackage{amssymb}
\usepackage{amsmath}
\usepackage{graphicx,color,dcolumn,booktabs,bm}
\usepackage{longtable,lscape}
\usepackage{txfonts}
\usepackage{overpic}
\usepackage{indentfirst}
\usepackage{cases}
\usepackage{multirow}
\usepackage{epstopdf}
\usepackage{ulem}
\usepackage[colorlinks,
            citecolor=blue,
            anchorcolor=red,
            menucolor=red,
            linkcolor=red,
            filecolor=red,
            runcolor=red,
            urlcolor=blue,
            frenchlinks=false]{hyperref}
\usepackage{threeparttable}
\allowdisplaybreaks
\begin{document}

 \title{Radiative decays of vector mesons with light-cone sum rules}
\author{Zhi Jun Wang$^{1}$}
\email{2023212251@nwnu.edu.cn}
\author{Di Gao$^{1,2,3}$}
\email{digao@impcas.ac.cn}
\author{Kai Kai Zhang$^{1}$}
\email{zhangkk314@outlook.com}
\author{Yuan Yuan Ma$^{1}$}
\email{Mayuanyuan917@163.com}
\author{Yan Jun Sun$^{1}$}\thanks{Corresponding author}
\email{sunyanjun@nwnu.edu.cn}
\affiliation{
	$^1$Institute of Theoretical Physics, College of Physics and
	Electronic Engineering, Northwest Normal University, Lanzhou 730070, China \\
	$^2$Institute of Modern Physics, Chinese Academy of Sciences, Lanzhou 730000, China\\
	$^3$Lebedev Physical Institute, Russian Academy of Sciences, Leninsky Prospekt 53, 119991, Moscow, Russia \\
	}
\date{\today}

 \begin{abstract}
Hadronic electromagnetic form factors and radiative decay properties offer a crucial window into the nonperturbative dynamics of quantum chromodynamics (QCD). In this work, we employ the light-cone sum rules (LCSR) method to systematically investigate the M1 radiative decay of vector mesons. Our study covers processes including $K^{*-}\rightarrow K^-\gamma$, $D^*\rightarrow D\gamma$, $B^*\rightarrow B\gamma$, $D^{*+}_s\rightarrow D^+_s\gamma$, and $B_s^*\rightarrow B_s\gamma$, and further extends to the excited charmonium state $\psi(2S)$. Our calculations yield decay widths for $K^*$ and $\psi(2S)$ that are in excellent agreement with experimental data. For the charm and bottom meson decays, where precise measurements are lacking, we provide theoretical predictions and compare them with other theoretical approaches. Most notably, our analysis reveals a universal linear dependence of the decay width on a function A(x) in the logarithmic coordinate system, which originates from the two-body decay dynamics and the ratio of the initial and final state decay constants. This relationship holds for the ground state $V \rightarrow P \gamma $ processes here and suggests a broader applicability to radiative decays of ground-state vector mesons.

\end{abstract}


\maketitle
\date{\today}

\section{INTRODUCTION}
\label{sec:in}
Quantum chromodynamics (QCD) is the fundamental theory describing the strong interaction. The property of asymptotic freedom enables the application of perturbative QCD in the high energy regime, leading to remarkable successes in describing both inclusive and exclusive processes at large momentum transfer.  Nevertheless, investigating the properties of QCD in the nonperturbative domain represents a central challenge in modern particle physics. Electromagnetic transitions provide a crucial window into this domain. Among these, magnetic dipole (M1) transitions, where a spin-1 vector meson decays into a spin-0 pseudoscalar meson via photon emission, serve as powerful probes for investigating hadron structure. The corresponding transition form factors are pivotal for deepening our insight into nonperturbative QCD dynamics \cite{Gao:2003ag,Pacetti:2014jai,Wilson:1974sk,Luscher:1996ug,Bali:2000gf,Tran:2025fhb,Belyaev:1993wp,Ridwan:2024ngc}. At present, the Particle Data Group (PDG) \cite{ParticleDataGroup:2024cfk} has compiled extensive experimental data on hadron radiative decays. This highlights the demand for robust theoretical frameworks capable of interpreting these results and predicting new decay channels.

Currently, M1 radiative decays have been experimentally observed for numerous mesons. For example, for the ground state magnetic dipole (M1) radiative transition $K^{*-}\to K^-\gamma$, the decay width is $50.37$ keV  \cite{ParticleDataGroup:2024cfk}. For the charm-containing processes $D^*\rightarrow D\gamma$ and $D^{*+}_s\rightarrow D^+_s\gamma$, only upper limits are available for  the M1 radiative decay widths specifically  $2.1$ MeV and $1.9$ MeV, respectively \cite{Asratian:1990bu,Abachi:1988fw,CLEO:1995ewe,ARGUS:1994bem}. Recently, the latest data from the Beijing Spectrometer III (BESIII) Collaboration reported measurements of the branching fractions for $D^*\rightarrow D\gamma$ and $D^{*+}_s\rightarrow D^+_s\gamma$ as $(34.5\pm0.8 \pm0.5)\%$ and $(93.57\pm0.38\pm0.22)\%$, respectively \cite{BESIII:2014rqs,BESIII:2022kbd}. However, although the M1 radiative decays of the $B^*$ and $B_s^*$ mesons have been experimentally observed, their decay widths still lack precise experimental measurements \cite{ParticleDataGroup:2024cfk}. For the excited state magnetic dipole radiative transition $\psi(2S) \rightarrow \eta_c(2S) \gamma $, this transition poses a particular challenge due to the narrow phase space of $\eta_c(2S)$. Subsequent studies have confirmed that $\eta_c(2S)$ can be produced via radiative decay of $\psi(2S)$. The Crystal Ball group \cite{Edwards:1982fif} reported an excess of $E_\gamma \simeq 91$ MeV gamma rays from inclusive $\psi(2S) \rightarrow X \gamma $ decays and interpreted this as possible evidence for the $\eta_c(2S)$ with mass $3594\pm 5$ MeV/$c^2$. The Belle Collaboration later observed the pseudoscalar meson $\eta_c(2S)$ in $B$ meson decays \cite{Belle:2002bnx}. Recently, the  BESIII Collaboration investigated the M1 transition in the decay $\psi(2S)\rightarrow\eta_c(2S)\gamma$, measuring the corresponding branching fraction as $\mathcal{B}(\psi(3686)\rightarrow\eta_c(2S)\gamma) =(5.2\pm0.3\pm0.5^{+1.9}_{-1.4}$)$\times10^{-4}$, and the decay width as $\Gamma(\psi(3686)\rightarrow\eta_c(2S)\gamma) = 0.15^{+0.06}_{-0.04}$ keV \cite{BESIII:2023lcc}. Theoretically, a wide array of theoretical frameworks has been developed for M1 radiative decays, such as Godfrey-Isgur model \cite{Godfrey:1985xj,Barnes:2005pb,Godfrey:2015dva}, relativistic quark model \cite{Goity:2000dk}, light-front quark model (LFQM) \cite{Peng:2012tr}, light-cone quark model \cite{Choi:1997iq}, meson loop corrections  \cite{Li:2007xr,Li:2011ssa}, QCD sum rules (QCDSR) \cite{Aliev:1994nq,Dosch:1995kw,Zhu:1996qy,Lu:2024tgy}, lattice quantum chromodynamics  \cite{Dudek:2009kk,Owen:2015gva,Meng:2024gpd,Colquhoun:2023zbc,Alexandrou:2018jbt,Lee:2008qf,Becirevic:2009xp}, and light-cone sum rules (LCSR) \cite{Guo:2019xqa,Pullin:2021ebn,Aliev:2009gj,Aliev:2019lsd}. Nevertheless, significant challenges persist in the study of radiative decays and related research areas, as exemplified by the $\rho - \pi$ puzzle \cite{Anselmino:1991es,Chao:1996sf,Wang:2021dxw} and several anomalous decay phenomena. Resolving these puzzles and advancing our fundamental understanding of hadronic structure and its underlying interaction mechanisms will necessitate sustained and concerted investigation on both experimental and theoretical fronts.

Light-cone sum rules (LCSR) is a nonperturbative method based on the first principles of QCD. It combines the operator product expansion (OPE) with a description of nonperturbative dynamics in terms of light-cone distribution amplitude (LCDA), and has been proven highly effective in calculating key nonperturbative parameters for exclusive processes \cite{Shifman:1978bx,Braun:1988qv}. The method exhibits a comprehensive scope of applicability, enabling calculations of transition form factors \cite{CerqueiraJr:2021afn,Zhao:2021sje,Peng:2019apl}, coupling constants \cite{Azizi:2010sy}, decay constants \cite{Lucha:2016nzv,Wang:2015mxa}, magnetic moments of multiquark states and other physical quantities      \cite{Wan:2024ykm,Agaev:2024wvp,Wang:2024ciy,Ozdem:2018uue,Ozdem:2022iqk}. While the LCSR method itself is not new, its systematic application to several vector-to-pseudoscalar M1 transitions including $\psi(2S)\rightarrow \eta_c(2S)\gamma$, supplements the existing literature. Unlike the photon wave functions employed in the literature \cite{Aliev:1995zlh}, we adopt the light-cone distribution amplitudes of pseudoscalar mesons as the nonperturbative inputs.

In this work, we employ LCSR to carry out a systematic study of the M1 radiative decays of vector mesons ($K^{*-}$, $B^*$, $B_s^*$, $D^*$, $D^{*+}_s$) and extend the formalism to the radiative transition of the excited charmonium state $\psi(2S)\rightarrow\eta_c(2S)\gamma$. In addition to computing the transition form factors and decay widths and comparing them with existing data and other theoretical approaches, we have performed a comprehensive analysis of the systematics of these decays. Notably we attempt to analyze the behavior of the form factor across different transition channels with respect to the mass scale and a universal linear relationship between the decay width and a specific kinematic function $A(x)$ is founded. This observed scaling may offer a fresh, unifying perspective on the systematics of radiative decays and could provide valuable insights into understanding broader hadronic anomalies, such as the $\rho-\pi$ puzzle, by revealing underlying universal dynamics in M1 transitions.

This work is organized as follows. Section \ref{sec:fTor} outlines the theoretical framework and details the derivation of the transition form factors within the light-cone sum rules. Section \ref{sec:fTToor} presents the numerical results, discusses their physical implications, and compares them with available experimental data and other theoretical approaches. Finally, Section \ref{sec:summary} summarizes our findings and provides an outlook for future research.

 \section{Form Factor with LCSR }
\label{sec:fTor}

\subsection{Form factor for $K^*$  transition
}
\label{sec:ci}
This section details the derivation of the transition form factors for the M1 radiative decays within the LCSR framework. We first present a comprehensive calculation for the process $K^{*-} \rightarrow K^- \gamma$. The formalism for other heavy-light meson transitions ($D^*$, $B^*$, $D_s^*$, $B_s^*$) follows an analogous procedure, with the final results summarized for brevity. Subsequently, we extend the formalism to the charmonium sector to address the excited state transition $\psi(2S) \rightarrow \eta_c(2S) \gamma$. For the $K^{*-}\rightarrow K^-\gamma$ decay channel, the following correlation function is employed

\begin{equation}\label{for:Ha}
	\begin{aligned}
		\Pi_{\mu\nu}(Q,q)=i\int d^4xe^{iq\cdot x}\langle K^-(Q)|T\{J_\mu(x)J_\nu(0)\}|0\rangle,
	\end{aligned}
\end{equation}
where $Q$ is the four-momentum of the final-state $K^-$ meson, and $q$ is the four-momentum of the outgoing photon. The currents are defined using two types of constructions: one with $J_\mu=\overline{s}(x)\gamma_\mu s(x)$ and $J_\nu=\overline{u}(x)\gamma_\nu s(x)$, and the other with $J_\mu=\overline{u}(x)\gamma_\mu u(x)$ and $J_\nu=\overline{u}(x)\gamma_\nu s(x)$. These correspond to the two ways in which M1 radiative decay occurs: the photon is emitted either from the $s$-quark line or from the $u$-quark line.

On the phenomenological side, a complete set of  hadronic states sharing the same quantum numbers as the final states is inserted between the two types of currents ( Here, one of the two types of constructions is taken as an example for illustration, and the other yields analogous results on the phenomenological side. ), yielding
\begin{equation}\label{for:HRQa}
	\begin{aligned}
		\Pi_{\mu\nu}(Q,q) &= i\int d^4x\, e^{iq\cdot x} \sum_{n} \int \frac{d^3\vec{P}_1}{(2\pi)^3 2E_1} \theta(x_0-0) e^{iQ\cdot x} e^{-iP_1\cdot x} \langle K^-(Q) | \bar{s}(0)\gamma_\mu s(0) | n \rangle \langle n | \bar{u}(0)\gamma_\nu s(0) | 0 \rangle \\
		&\quad + i\int d^4x\, e^{iq\cdot x} \sum_{n} \int \frac{d^3\vec{P}_1}{(2\pi)^3 2E_1} \theta(0-x_0) e^{iP_1\cdot x} \langle K^-(Q) | \bar{u}(0)\gamma_\nu s(0) | n \rangle \langle n | \bar{s}(0)\gamma_\mu s(0) | 0 \rangle.
	\end{aligned}
\end{equation}
Isolating the ground state contributions gives
\begin{equation}\label{aaWa}
	\begin{aligned}
	\Pi_{\mu\nu}(Q,q)&=\frac{1}{m^2_{K^{*-}}-(q+Q)^2}\langle K^-(Q)|\bar{s}(0)\gamma_\mu s(0)|K^{*-}\rangle\langle K^{*-}|\bar{u}(0)\gamma_\nu s(0)|0\rangle\\
	&\quad
	+\int_{s_0}^{\infty}ds\frac{\rho(Q^2,s)}{s-(Q+q)^2},
	\end{aligned}
\end{equation}
where $s_0$ is the threshold parameter that separates the ground state from excited states. Following Ref. \cite{Dudek:2006ej}, the matrix elements are parametrized as

\begin{equation}\label{qaqa}
	\begin{aligned}
		\langle K^-(Q)|\overline{s}(0)\gamma_\mu s(0)|K^{*-}\rangle=\varepsilon_{\mu\alpha\beta\gamma} Q^\beta q^\alpha\varepsilon^{*\gamma}F_{VP}\frac{2}{m_{K^{*-}}+m_{K^-}},
	\end{aligned}
\end{equation}
\begin{equation}\label{for:deqa}
	\begin{aligned}
		\langle K^{*-}|\overline{u}(0)\gamma_\nu s(0)|0\rangle=m_{K^{*-}}f_{K^{*-}}\varepsilon_\nu,
	\end{aligned}
\end{equation}
where, $m_{K^{*-}}$ is the mass of the $K^{*-}$ meson, $m_{K^-}$ is the mass of the $K^-$ meson, $ f_{K^{*-}} $ is the decay constant, $ \varepsilon_\nu $ is the polarization vector,  $\varepsilon_{\mu\alpha\beta\gamma}$ is a totally antisymmetric tensor and $ F_{VP} $ is the transition form factor. With these parametrizations, the correlator on the phenomenological side becomes

\begin{equation}\label{for:exw}
	\begin{aligned}
		\Pi_{\mu\nu}(Q,q)&=\frac{1}{m_{K^{*-}}^2-(Q+q)^2}\frac{2\varepsilon_{\mu\alpha\beta\gamma}Q^\beta q^\alpha\varepsilon^{*\gamma}F_{VP}m_{ K^{*-}} f_{K^{*-}}\varepsilon_\nu}{m_{K^{*-}}+m_{K^-}}\\
		&\quad
		+\int_{s_0}^{\infty}ds\frac{\rho(Q^2,s)}{s-(Q+q)^2}.		
	\end{aligned}
\end{equation}

On the theoretical side of the correlator, the operator product expansion (OPE) is employed. After contracting the quark fields for the two types of currents, the corresponding propagators are substituted accordingly.
By substituting the $s$-quark propagator $S(0,x)=i\int\frac{d^4k}{(2\pi)^4}e^{-ik \cdot x}\{-\frac{m_s-k^{\alpha}\gamma_{\alpha}}{m^2_s-k^2}\}$ and the $u$-quark propagator $S(x,0)=i\int\frac{d^4k}{(2\pi)^4}e^{-ik \cdot x}\{-\frac{m_u+k^{\alpha}\gamma_{\alpha}}{m^2_u-k^2}\}$into the correlation function, we obtain, corresponding to two types of current constructions

\begin{equation}\label{for:mu}
	\begin{aligned}
	&\	\Pi_{\mu\nu}(Q,q)=-\int d^4xe^{iqx}\int\frac{d^4k}{(2\pi)^4}e^{-ik\cdot x}\frac{k^\alpha}{m_s^2-k^2}\langle K^-(Q)|\overline{u}(0)\gamma_\nu\gamma_\alpha\gamma_\mu s(x)|0\rangle, \\
&\ \Pi_{\mu\nu}(Q,q)=\int d^4xe^{iqx}\int\frac{d^4k}{(2\pi)^4}e^{-ik\cdot x}\frac{k^\alpha}{m_u^2-k^2}\langle K^-(Q)|\overline{u}(x)\gamma_\mu\gamma_\alpha\gamma_\nu s(0)|0\rangle.
	\end{aligned}
\end{equation}
\noindent

The matrix elements in the above expression can be simplified using the following formula: $\gamma_{\mu}\gamma_{\alpha}\gamma_{\nu}=S_{\mu\alpha\nu\beta}\gamma^{\beta}-i\varepsilon_{\mu\alpha\nu\beta}\gamma^{\beta}\gamma_5$.

Since the light-cone distribution amplitude of $K$ meson is defined as \cite{Sun:2010oyk}
\begin{equation}\label{for:wavee}
	\begin{aligned}
		&\ \langle K^-(Q)|\overline{u}(0)\gamma^\beta\gamma_5s(x)|0\rangle=-iQ^\beta f_{K^-}\int_0^1 due^{i(1-u)Q\cdot x}\phi_{K^-}(u)+higher\enspace twist\enspace terms, \\
&\ \langle K^-(Q)|\overline{u}(x)\gamma^\beta\gamma_5s(0)|0\rangle=-iQ^\beta f_{K^-}\int_0^1 due^{iuQ\cdot x}\phi_{K^-}(u)+higher\enspace twist\enspace terms.
	\end{aligned}
\end{equation}
The above expression provides the parameterization of the matrix element of the K meson near the light cone. This matrix element can be described in terms of the decay constant $f_{K^-}$ and the light-cone distribution amplitudes $\phi_{K^-}(u)$. Here, $u$ denotes the momentum fraction carried by the $u$ quark, while $\bar{u}=1-u$ represents the momentum fraction carried by the $\bar{s}$ antiquark. The correlation function is expressed in the following Eq. (\ref{for:X}), corresponding to two types of constructions, respectively.

\begin{equation}\label{for:X}
	\begin{aligned}
		 &\  \Pi_{\mu\nu}(Q,q)=-\varepsilon_{\nu\alpha\mu\beta}f_{K^-}Q^\beta q^\alpha\int_0^1du\frac{\phi_{K^-}(u)}{m_s^2-(q+\bar{u}Q)^2}  ,\\
&\ \Pi_{\mu\nu}(Q,q)=-\varepsilon_{\mu\alpha\nu\beta}f_{K^-}Q^\beta q^\alpha\int_0^1du\frac{\phi_{K^-}(u)}{m_u^2-(q+uQ)^2}.
	\end{aligned}
\end{equation}

Applying the ``quark-hadron duality" assumption to the hadronic spectral density in Eq. (\ref{for:exw}),
\begin{equation}\label{for:T}
	\begin{aligned}
		\int_{0}^\infty ds\frac{\rho^h(s)}{s-q^2}\simeq\frac{1}{\pi}\int_{0}^\infty ds \frac{Im\Pi^{pert}(s)}{s-q^2},
	\end{aligned}
\end{equation}
and matching the theoretical side with the phenomenological side yields

\begin{equation}\label{DN}
	\begin{aligned}
	 \frac{F_{VP}}{m^2_{K^{*-}}-(q+Q)^2}=\frac{1}{2}\frac{f_{K^-}}{f_{K^{*-}}}\frac{m_{K^{*-}}+m_{K^-}}{m_{K^*-}}\left[\int_0^{\Delta_1}du\frac{\phi_{K^-}(u)}{m^2_s-(q+\bar{u}Q)^2}+
\int_{\Delta_2}^1du\frac{\phi_{K^-}(u)}{m^2_u-(q+uQ)^2}\right],
	\end{aligned}
\end{equation}
where the upper limit $\Delta_1$  and the lower limit $\Delta_2$ are given respectively as
\begin{equation}\label{Delta}
	\begin{aligned}
	& \Delta_1=\frac{1}{2m^2_{K^-}}\left[-\sqrt{(s_0+m_{K^-}^2+P^{\prime2})^2-4m_{K^-}^2(s_0-m^2_s)}+(s_0+m_{K^-}^2-q^2)\right] , &\\ & \Delta_2=\frac{1}{2m^2_{K^-}}\left[\sqrt{(s_0-m_{K^-}^2+P^{\prime2})^2+4m_{K^-}^2(m_u^2+P^{\prime2})}-(s_0-m_{K^-}^2-q^2)\right] ,
	\end{aligned}
\end{equation}
where the four-momenta are specified as $Q^2=m_{K^-}^2$ and $P^{\prime2}=-q^2$.

To suppress the contributions from higher excited states and the continuum, we apply the Borel transformation \cite{Shifman:1978bx,Sun:2010oyk,Lepage:1980fj} to the four-momenta $(Q+q)$ , $(uQ+q)$ and $(\bar{u}Q+q)$,
\begin{equation}\label{for:ff}
	\begin{aligned}
	&\mathcal{B}_{M^2}\left[\frac{1}{m^2_{K^{*-}}-(Q+q)^2}\right] = \frac{1}{M^2} e^{-m^2_{K^{*-}}/M^2},&\\		
&\mathcal{B}_{M^2}\left[\frac{1}{m^2_s-(uQ+q)^2}\right]=\frac{1}{uM^2}e^{-\frac{1}{uM^2}[m^2_s+u(1-u)Q^2-(1-u)q^2]},&\\
&\mathcal{B}_{M^2}\left[\frac{1}{m^2_s-(\bar{u}Q+q)^2}\right]=\frac{1}{\bar{u}M^2}e^{-\frac{1}{\bar{u}M^2}[m^2_s+u(1-u)Q^2-uq^2]}.
	\end{aligned}
\end{equation}
Finally, the transition form factor is obtained

\begin{equation}\label{ffqf}
	\begin{aligned}
&F_{VP}=\frac{1}{2}\frac{f_{K^-}}{f_{K^{*-}}}\frac{m_{K^{*-}}+m_{K^-}}{m_{K^*-}} \left[\int^{\Delta_1}_0 du \frac{\phi_{K^-}(u)}{\bar{u}} e^{-\frac{1}{\bar{u}M^2}(m^2_s+u\bar{u}m^2_{K^-}-uq^2)+\frac{m^2_{K^{*-}}}{M^2}}
+ \int^{1}_{\Delta_2} du \frac{\phi_{K^-}(u)}{u} e^{-\frac{1}{uM^2}(m^2_u+u\bar{u}m^2_{K^-}-(1-u)q^2)+\frac{m^2_{K^{*-}}}{M^2}}\right].
	\end{aligned}
\end{equation}

Through phenomenological and theoretical treatments of the correlation function, followed by matching procedures, the form factor for the $K^{*-}\rightarrow K^-\gamma$ process has been obtained. This methodology is universally applicable to other processes, and for $D^*\rightarrow D\gamma$, $B^*\rightarrow B\gamma$, $D^{*+}_s\rightarrow D^+_s\gamma$ and $B_s^*\rightarrow B_s\gamma$, the form factors are

\begin{equation}
	\begin{aligned}
&F_{VP}=\frac{1}{2}\frac{f_D}{f_{D^*}}\frac{m_{D^*}+m_D}{m_{D^*}} \left[\int^{\Delta_1}_0 du \frac{\phi_D(u)}{\bar{u}} e^{-\frac{1}{\bar{u}M^2}(m^2_c+u\bar{u}m^2_{D}-uq^2)+\frac{m^2_{D^*}}{M^2}}+\int^{1}_{\Delta_2} du \frac{\phi_D(u)}{u} e^{-\frac{1}{uM^2}(m^2_u+u\bar{u}m^2_{D}-(1-u)q^2)+\frac{m^2_{D^*}}{M^2}}\right],\\
&F_{VP}=\frac{1}{2}\frac{f_{D_s^+}}{f_{D^{*+}_s}}\frac{m_{D^{*+}_s}+m_{D_s^+}}{m_{D^{*+}_s}}\left[\int^{\Delta_1}_0du\frac{\phi_{D_s^+}(u)}{\bar{u}}e^{-\frac{1}{\bar{u}M^2}(m^2_c+u\bar{u}m^2_{D_s^+}-uq^2)+\frac{m^2_{D^{*+}_s}}{M^2}}+\int^{1}_{\Delta_2}du\frac{\phi_{D_s^+}(u)}{u}e^{-\frac{1}{uM^2}(m^2_s+u\bar{u}m^2_{D_s^+}-(1-u)q^2)+\frac{m^2_{D^{*+}_s}}{M^2}}\right],\\	&F_{VP} =\frac{1}{2}\frac{f_B}{f_{B^*}}\frac{m_{B^*}+m_B}{m_{B^*}} \left[\int^{\Delta_1}_0 du \frac{\phi_B(u)}{\bar{u}} e^{-\frac{1}{\bar{u}M^2}(m^2_b+u\bar{u}m^2_{B}-uq^2)+\frac{m^2_{B^*}}{M^2}}+\int^{1}_{\Delta_2} du \frac{\phi_B(u)}{u} e^{-\frac{1}{uM^2}(m^2_d+u\bar{u}m^2_{B}-(1-u)q^2)+\frac{m^2_{B^*}}{M^2}}\right],\\
&F_{VP} =\frac{1}{2}\frac{f_{B_s}}{f_{B^*_s}}\frac{m_{B^*_s}+m_{B_s}}{m_{B^*_s}} \left[\int^{\Delta_1}_0 du \frac{\phi_{B_s}(u)}{\bar{u}}e^{-\frac{1}{\bar{u}M^2}(m^2_b+u\bar{u}m^2_{B_s}-uq^2)+\frac{m^2_{B^*_s}}{M^2}}+\int^{1}_{\Delta_2} du \frac{\phi_{B_s}(u)}{u} e^{-\frac{1}{uM^2}(m^2_s+u\bar{u}m^2_{B_s}-(1-u)q^2)+\frac{m^2_{B^*_s}}{M^2}}\right],\\
	\end{aligned}
\end{equation}
where $M^2$ is Borel parameter, $m_{D^*}$ ($m_{B^*}$, $m_{D^{*+}_s}$, $m_{B^*_s}$) is the mass of the initial state meson, $m_{D}$ ($m_{B}$, $m_{D^{+}_s}$, $m_{B_s}$) is the mass of the final state meson, and $f_{D^*}$ ($f_{B^*}$, $f_{D^{*+}_s}$, $f_{B^*_s}$) and $f_{D}$ ( $f_{B}$, $f_{D_s^+}$, $f_{B_s}$) are the decay constants of the initial and final state mesons, respectively. Additionally, $m_c$ and $m_b$ are the masses of the c quark and the b quark, respectively.

\subsection{Form factor for $\psi(2S)$ transition}

The LCSR method can also be applied to transitions involving excited states. We consider the M1 radiative decay \(\psi(2S) \to \eta_c(2S) \gamma\). The correlation function is defined as

\begin{equation}\label{for:BQQ}
	\begin{aligned}
	\Pi_{\mu\nu}(Q,q)=i\int d^4x e^{iq\cdot x}\langle \eta_c(2S)(Q)|T\{J_\mu^c(x)J_\nu^c(0)\}|0\rangle,
	\end{aligned}
\end{equation}
where the electromagnetic currents of the c quark are given by $J_\mu^c=\overline{c}(x)\gamma_\mu c(x)$ and  $J_\nu^c=\overline{c}(x)\gamma_\nu c(x)$.

On the phenomenological side of the correlation function, we insert the hadron state $\sum_{n}|n\rangle \langle n|$, which includes $J/\psi$, $\psi (2S)$, higher excited states, and the hadronic continuum. Since the mass of state $J/\psi$ is lower than that of state $\eta_c(2S)$, the transition $J/\psi \rightarrow \eta_c(2S)\gamma  $ cannot occur. Therefore, the transition $\psi(2S)\rightarrow\eta_c(2S)\gamma$  is isolated,

\begin{equation}\label{for:BQQTU}
	\begin{aligned}
		\Pi_{\mu\nu}(Q,q)&=\frac{1}{m^2_{\psi(2S)}-(q+Q)^2}\langle \eta_c(2S)(Q)|\bar{c}(0)\gamma_\mu c(0)|\psi(2S)\rangle\langle \psi(2S)|\bar{c}(0)\gamma_\nu c(0)|0\rangle\\
		&\quad
		+\int_{s_0}^{\infty}ds\frac{\rho(Q^2,s)}{s-(Q+q)^2},
	\end{aligned}
\end{equation}
where $\rho(Q^2,s)$ is defined similarly to Eq. (\ref{aaWa}).

The decay matrix elements in Eq. (\ref{for:BQQTU}) are parametrized as follows:

\begin{equation}\label{qaqaVV}
	\begin{aligned}
		\langle \eta_c(2S)(Q)|\overline{c}(0)\gamma_\mu c(0)|\psi(2S)\rangle=\varepsilon_{\mu\alpha\beta\gamma} Q^\beta q^\alpha\varepsilon^{\gamma*}F_{VP}\frac{2}{m_{\psi(2S)}+m_{\eta_c(2S)}},
	\end{aligned}
\end{equation}
\begin{equation}\label{for:deWEqa}
	\begin{aligned}
		\langle \psi(2S)|\overline{c}(0)\gamma_\nu c(0)|0\rangle=m_{\psi(2S)}f_{\psi(2S)}\varepsilon_\nu,
	\end{aligned}
\end{equation}
where $m_{\psi(2S)}$ is the mass of the $\psi(2S)$ meson, $m_{\eta_c(2S)}$ is the mass of the $\eta_c(2S)$ meson, $ f_{\psi(2S)} $ is the decay constant, and $ F_{VP} $ is the transition form factor. Through the substitution of Eq. (\ref{qaqaVV}) and (\ref{for:deWEqa}) into Eq. (\ref{for:BQQTU}), the phenomenological result is obtained
\begin{equation}\label{for:BQQqqQ}
	\begin{aligned}
		\Pi_{\mu\nu}(Q,q)&=\frac{1}{m^2_{\psi(2S)}-(Q+q)^2}\frac{2\varepsilon_{\mu\alpha\beta\gamma}Q^\beta q^\alpha \varepsilon^{\gamma *}F_{VP}m_{\psi(2S)}f_{\psi(2S)}\varepsilon_\nu}{m_{\psi(2S)}+m_{\eta_c(2S)}}\\
		&\quad
		+\int_{s_0}^{\infty}ds\frac{\frac{1}{\pi}Im\Pi(Q,q)\theta(s-s_0)}{s-(Q+q)^2}.
	\end{aligned}
\end{equation}

On the other hand, the theoretical result for the correlation function is given by
\begin{equation}\label{for:BQQ}
	\begin{aligned}
		\Pi_{\mu\nu}(Q,q)=2\varepsilon_{\mu\alpha\nu\beta}f_{\eta_c(2S)}Q^\beta q^\alpha\int_{0}^{1}du\frac{\phi_{\eta_c(2S)}(u)}{m_c^2-(q+uQ)^2},
	\end{aligned}
\end{equation}
where  $\phi_{\eta_c(2S)}(u)$  is defined analogously to Eq. (\ref{for:wavee}). Matching the phenomenological side with the theoretical side and  applying the Borel transformation to both sides, the analytical expression of the form factor for the process $\psi(2S)\rightarrow\eta_c(2S)\gamma$ is obtained
\begin{equation}\label{for:BQQ}
	\begin{aligned}
		F_{VP}=\frac{m_{\psi(2S)}+m_{\eta_c(2S)}}{m_{\psi(2S)}}\frac{f_{\eta_c(2S)}}{f_{\psi(2S)}}\int_{\Delta}^{1}du\frac{\phi_{\eta_c(2S)}(u)}{u}e^{-\frac{1}{uM^2}[m^2_c+u(1-u)m^2_{\eta_c(2S)}+(1-u)P^{\prime2}]+\frac{m^2_{\psi(2S)}}{M^2}},
	\end{aligned}
\end{equation}
where $\Delta=\frac{1}{2m^2_{\eta_c(2S)}}\left[\sqrt{(s_0-m_{\eta_c(2S)}^2+P^{\prime2})^2+4m_{\eta_c(2S)}^2(m^2_c+P^{\prime2})}-(s_0-m_{\eta_c(2S)}^2-q^2)\right]$.

\subsection{Light-Cone distribution amplitudes}

The LCDAs serve as essential nonperturbative input parameters. The typical functional forms of LCDAs employed in this work, along with the required parameters, are provided below.

For the process $K^{*-}\rightarrow K^-\gamma$
, the LCDA of $K^-$ meson is chosen as \cite{Khodjamirian:2000ds}
\begin{equation}\label{for:BQQ}
	\begin{aligned}
		\phi_K(u,\mu)=6u(1-u)[1+\sum_{n=1}^{4}a_n^K(\mu)C_n^{3/2}(2u-1)],
	\end{aligned}
\end{equation}
where $C^{3/2}_n$ denotes the Gegenbauer polynomials and $a_n^K$ represents the corresponding Gegenbauer moments. In our calculation, the adopted Gegenbauer coefficients are $a_1^K = 0.09 \pm 0.13$ and $a_2^K = 0.03 \pm 0.03$ \cite{Choi:2007yua}.

For the decays $B^*\rightarrow B\gamma$, $B_s^*\rightarrow B_s\gamma$, \textcolor{blue}{$ D^*\rightarrow D\gamma $} and $D_s^{*+}\rightarrow D_s^+\gamma$, the LCDAs of the heavy-light mesons $B$, $D$, $D_s^+$ and $B_s$ are parametrized exponentially as \cite{Serna:2020txe}

\begin{equation}\label{for:BQQQ}
	\begin{aligned}
		\phi_H(u,\mu)=N(\alpha,\beta)4u\bar{u}e^{4\alpha u\bar{u}+\beta(u-\bar{u})},
	\end{aligned}
\end{equation}
where
\begin{equation}\label{for:BQQQ}
	\begin{aligned}
		N(\alpha,\beta)=16\alpha^{5/2}[4\sqrt{\alpha}(\beta sinh(\beta)+2\alpha cosh(\beta))+e^{\alpha+\frac{\beta^2}{4\alpha}}(-2\alpha+4\alpha^2-\beta^2)\times \left\{Erf(\frac{\alpha-\beta}{2\sqrt{\alpha}})+Erf(\frac{\alpha+\beta}{2\sqrt{\alpha}})\right\}]^{-1}.
	\end{aligned}
\end{equation}

The parameters $\alpha$ and $\beta$ are presented in Table \ref{tab:paraaaa}, and Erf here denotes the error function.

\begin{table}[h!]
	\caption{The parameters $\alpha$ and $\beta$ in the LCDAs of $D$, $B$, $D_s^+$ and $B_s$ mesons at the scale $\mu=2$ GeV \cite{Serna:2020txe}.}
	\centering
	\label{tab:paraaaa}
	\begin{tabular}{lcc}
		\bottomrule[1.0pt]\bottomrule[0.5pt]
		& $\alpha$       &$\beta$ \\\hline
		$ D^*\rightarrow D\gamma $&0.038$\pm$0.005 &1.431$\pm$0.085 \\
		$B^*\rightarrow B\gamma$ &  0.360$\pm$0.017     &5.706$\pm$0.225\\
		$D^{*+}_s\rightarrow D^+_s\gamma$ &  0.712$\pm$0.157       &$0.929\pm0.082$ \\
		$B_s^*\rightarrow B_s\gamma$ & $1.205\pm0.526$     &6.109$\pm$0.594  \\

		\bottomrule[0.5pt]\bottomrule[1.0pt]
	\end{tabular}
\end{table}

For the process $\psi(2S) \rightarrow \eta_c(2S)\gamma$, the BHL approach \cite{Lepage:1980fj,Huang:2007kb,Huang:1994dy} is used to define the LCDA of  $\eta_c(2S)$ meson,

\begin{equation}\label{for:VBQQQQ}
	\begin{aligned}
		\phi_{\eta_c(2S)}(u, \mu_0)&=\frac{2\sqrt{6}}{f_{\eta_c(2S)}}\int_{|\vec{k}_\perp|^2<\mu_0^2}^{}\frac{d^2\vec{k}_\perp}{16\pi^3}\Psi_{\eta_c(2S)}(u,\vec{k}_\perp),
	\end{aligned}
\end{equation}

\begin{equation}\label{for:jVBQQQQ}
	\begin{aligned}
		\Psi_{\eta_c(2S)}^{\lambda_1\lambda_2}(u,\vec{k}_\perp)&=\phi_{BHL}(u, \vec{k}_\perp)\chi^{\lambda_1\lambda_2}(u, \vec{k}_\perp)=Ae^{-b^2\frac{\vec{k}_\perp^2+m_c^{*2}}{u(1-u)}}\chi^{\lambda_1\lambda_2}(u,\vec{k}_\perp),
	\end{aligned}
\end{equation}
where $\Psi_{\eta_c(2S)}^{\lambda_1\lambda_2}(u,\vec{k}_\perp)$ is the wave function of $\eta_c(2S)$
meson, and $\chi^{\lambda_1\lambda_2}(u, \vec{k}_\perp)$ \cite{Cao:1997hw,Huang:2004su,Huang:2006wt} is the spin wave function, as presented in Table \ref{tab:parass}. The parameters $A$ and $b^2$ are determined by the following two constraints:

\begin{equation}\label{for:VVQBQQQ}
	\begin{aligned}
		&\frac{2\sqrt{6}}{f_{\eta_c(2S)}}\int_{0}^{1}dx\int_{|\vec{k}_\perp|^2<\mu_0^2}\frac{d^2\vec{k}_\perp}{16\pi^3}\sum_{\lambda_1\lambda_2}\Psi_{\eta_c(2S)}^{\lambda_1\lambda_2}(u_i,\vec{k}_\perp)=1,\\
		&\int_{0}^{1}dx\int\frac{d^2\vec{k}_\perp}{16\pi^3}|\phi_{BHL}(u,\vec{k}_\perp)|^2=P_{\eta_c(2S)}.
	\end{aligned}
\end{equation}
The first equation in (\ref{for:VVQBQQQ}) is the normalization condition, and the second equation corresponds to the probability normalization condition, where $P_{\eta_c(2S)}$ denotes the proportion of the Fock state. Here, we select the leading order in the Fock state expansion with $P_{\eta_c(2S)} = 0.5$, and assume that the initial scale of the distribution amplitude of $\eta_c(2S)$ meson is $\mu_0=m^*_c$. Using these inputs, the two undetermined parameters are obtained as $A =21.7713$ GeV$^{-1}$ and $b^2 = 0.06787$ GeV$^{-2}$.

\begin{table}[h!]
	\caption{The spin wave function for $\eta_c(2S)$. Here, $m_c^*=1.8$ GeV \cite{Bondar:2004sv} is the constituent quark mass of the c quark, and $\lambda_1$ and $\lambda_2$ denote the helicities of the c quark and the anti-c quark, respectively.  }
	\label{tab:parass}
	\begin{tabular}{lcccc}
		\bottomrule[1.0pt]\bottomrule[0.5pt]
		& $\uparrow\uparrow$       &$\uparrow\downarrow$       &$\downarrow\uparrow$  &$\downarrow\downarrow$ \\\hline
		
		$\chi^{\lambda_1\lambda_2}(u,\vec{k}_\perp)$ & $-\frac{k_x-ik_y}{\sqrt{2(m_c^{*2}+k^2_{\perp})}}$  &$\frac{m_c^*}{\sqrt{2(m_c^{*2}+k^2_{\perp})}}$ &-$\frac{m_c^*}{\sqrt{2(m_c^{*2}+k^2_{\perp})}}$ & -$\frac{k_x+ik_y}{\sqrt{2(m_c^{*2}+k^2_{\perp})}}$\\
		
		\bottomrule[0.5pt]\bottomrule[1.0pt]
	\end{tabular}
\end{table}

\begin{figure}[H]
    \centering
    \vspace{-0.50cm}
    \subfigure{
        \includegraphics[width=1.00\linewidth]{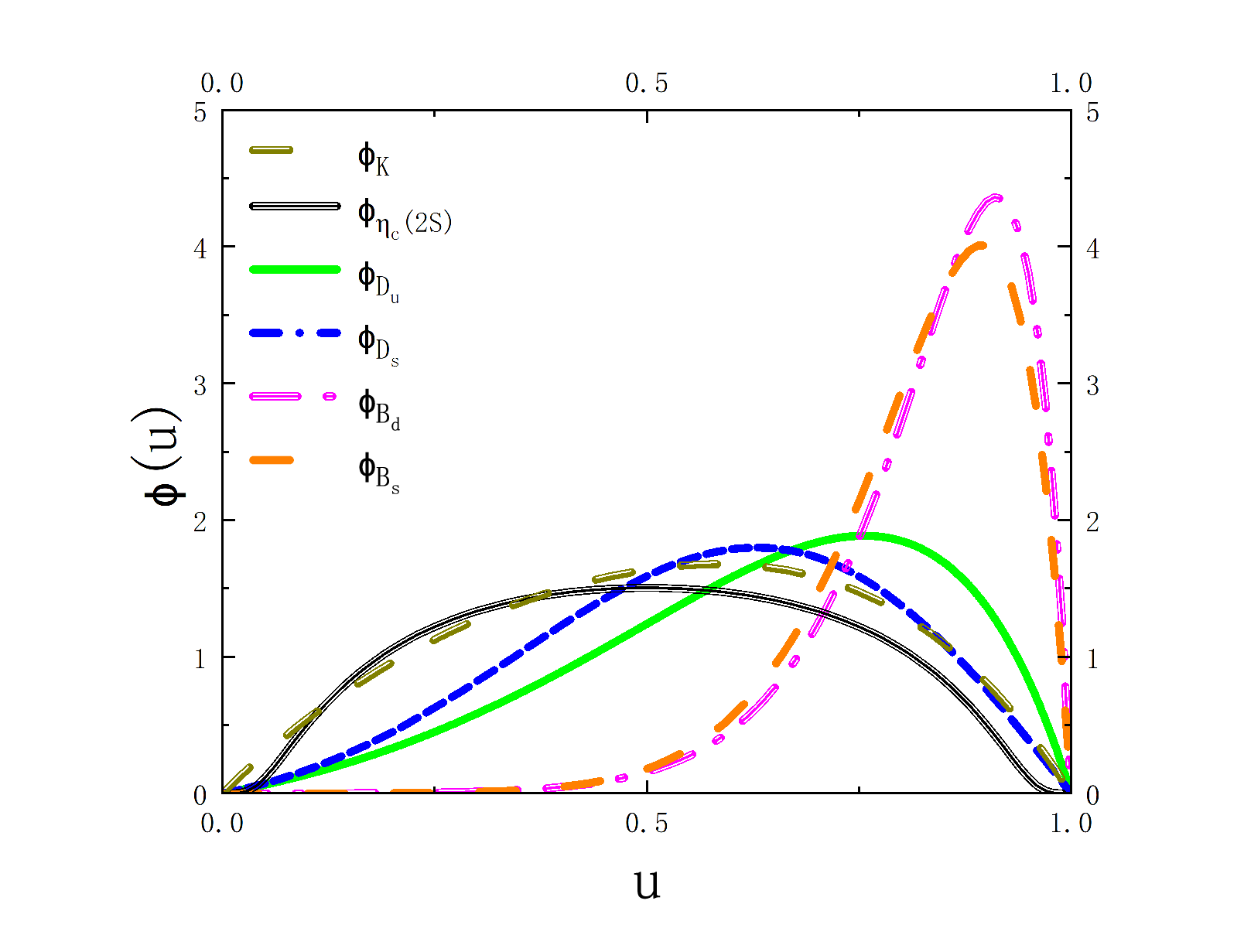}
        \label{6gfb.eps}
    }
    \caption{The distribution amplitudes of the $K$, $\eta_c(2S)$, $D$, $D_s$, $B$, and $B_s$ mesons.}
    \label{6gfb}
\end{figure}

In the above, the distribution amplitudes of each process were determined. Below, the images of the selected distribution amplitudes are presented as shown in Fig. \ref{6gfb}, from which it is evident that the distribution amplitude of $\eta_c(2S)$ has no nodes.

\subsection{Decay width}

Once the form factors are determined, the M1 radiative decay widths for the processes $K^{*-}\rightarrow K^-\gamma$, $D^*\rightarrow D\gamma$, $B^*\rightarrow B\gamma$, $D^{*+}_s\rightarrow D^+_s\gamma$, and $B_s^*\rightarrow B_s\gamma$, can be obtained \cite{Choi:2007se,Dudek:2006ej}

\begin{equation}\label{for:OBQQQ}
	\begin{aligned}
		\Gamma=\frac{\alpha}{3} \frac{|F_{VP}|^2}{(m_i+m_f)^2} k^3_\gamma,
	\end{aligned}
\end{equation}
where $\alpha$ is the fine structure constant, $k_\gamma=(M^2_i-M^2_f)/2M_i$, and the coupling constant $F_{VP}(q^2)$ for a real photon is equal to the form factor $F_{VP}(q^2)$ in the limit $q^2\rightarrow 0$.

\section{Numerical Results and Discussion}
\label{sec:fTToor}

\subsection{Input parameters }

In the preceding sections, analytic expressions for the form factors and decay widths of the six decay processes have been derived. In the subsequent discussion, numerical results will be provided and compared with experimental measurements along with predictions from other theoretical approaches.

The input parameters consist of two parts: some are taken from the PDG \cite{ParticleDataGroup:2024cfk} (e.g., quark masses), compiled in Table \ref{tab:parad} and others are required by the LCSR analysis, including the thresholds and the Borel parameters.

\renewcommand{\tabcolsep}{0.1cm}
\renewcommand{\arraystretch}{1.0}
\begin{table}[h!]
	\caption{Parameters required for all processes include quark masses, decay constants, and meson masses. }
	\label{tab:parad}
	\begin{tabular}{lccc}
		\bottomrule[1.0pt]\bottomrule[0.5pt]
		Processes & $Quark$ $masses$(GeV) \cite{ParticleDataGroup:2024cfk}&	$Decay$ $constants$(GeV) \cite{Arifi:2022pal,Wang:2002zba} & $Meson$ $masses$(GeV) \cite{ParticleDataGroup:2024cfk}\\\hline
		
		$K^{*-}\rightarrow K^-\gamma$ &  $m_s=0.0935\pm0.0005$, $m_u=0.00216\pm0.00004$     &$f_{K^-}=0.160$ , $f_{K^{*-}}=0.226\pm0.028$&      $m_{K^-}=0.4936$ , $m_{K^{*-}}=0.890$\\
		$D^*\rightarrow D\gamma$ &  $m_c=1.2730\pm0.0028$, $m_u=0.00216\pm0.00004$&$f_D=0.2067\pm8.9$ , $f_D^*=0.265$&$m_D=1.864$ , $m_{D^*}=2.007$ \\
		$B^*\rightarrow B\gamma$  & $m_b=4.183\pm0.004$, $m_d=0.0047\pm0.00004$  &$f_B=0.188\pm0.025$ , $f_B^*=0.215$  &$m_B=5.279$ , $m_{B^*}=5.324$\\
		$D^{*+}_s\rightarrow D^+_s\gamma$ &$m_c=1.2730\pm0.0028$, $m_s=0.0935\pm0.0005$  &$f_{D^+_s}=0.219$ , $f_{D^{*+}_s}=0.375\pm0.024$&$m_{D^+_s}=1.968$ , $m_{D_s^{*+}}=2.106$\\
		$B_s^*\rightarrow B_s\gamma$ &$m_b=4.183\pm0.004$, $m_s=0.0935\pm0.0005$ &$f_{B_s}=0.235$ , $f_{B^*_s}=0.256$
		&$m_{B_s}=5.366$ , $m_{B^*_s}=5.415$\\
		$\psi(2S)\rightarrow \eta _c(2S)\gamma$ &$m_c=1.2730\pm0.0028$&$f_{\eta_c(2S)}=0.318,f_{\psi(2S)}=0.294$&$m_{\eta_c(2S)}=3.637,m_{\psi (2S)}=3.686$\\
		\bottomrule[0.5pt]\bottomrule[1.0pt]
	\end{tabular}
\end{table}

Within the LCSR framework, the threshold and the Borel parameter are crucial input parameters. As a common practice, the threshold is taken to be the squared mass of the first excited state meson, although its precise value depends on the specific decay channel and must be fixed separately. For the decay $K^{*-} \rightarrow K^-\gamma$, the first excited state of the $K^*$ meson, $K^*(1410)$, has a mass of 1.414 GeV, so the threshold $s_0$ is approximately $1.9 \, \text{GeV}^2$. The Borel parameter is typically determined by imposing two criteria:
\begin{enumerate}
\item The combined contribution from higher resonances and the continuum is less than 30\%;
\item The physical observables exhibit only mild dependence on the Borel parameter.
\end{enumerate}

\subsection{Transition form factors}
\label{sec:bc}

Based on the criteria and constraints established above, the dependence of the form factors on the threshold $s_0$ and Borel parameter $M^2$ is discussed as follows.

\begin{figure}[H]
    \centering
    \vspace{-0.50cm}
    \subfigure[]{
        \includegraphics[width=0.45\linewidth]{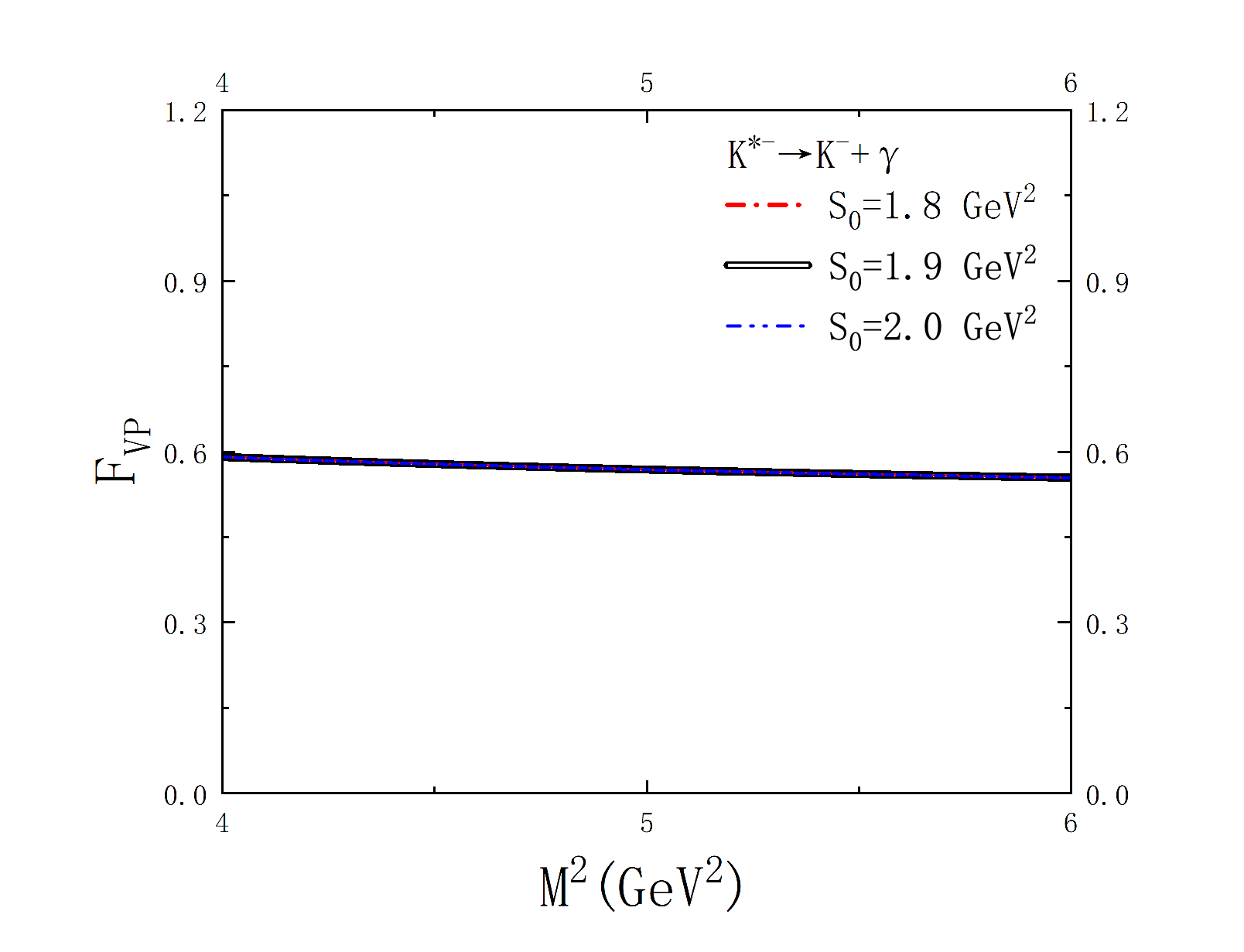}
        \label{Kf.eps}
    }
    \quad
    \subfigure[]{
        \includegraphics[width=0.45\linewidth]{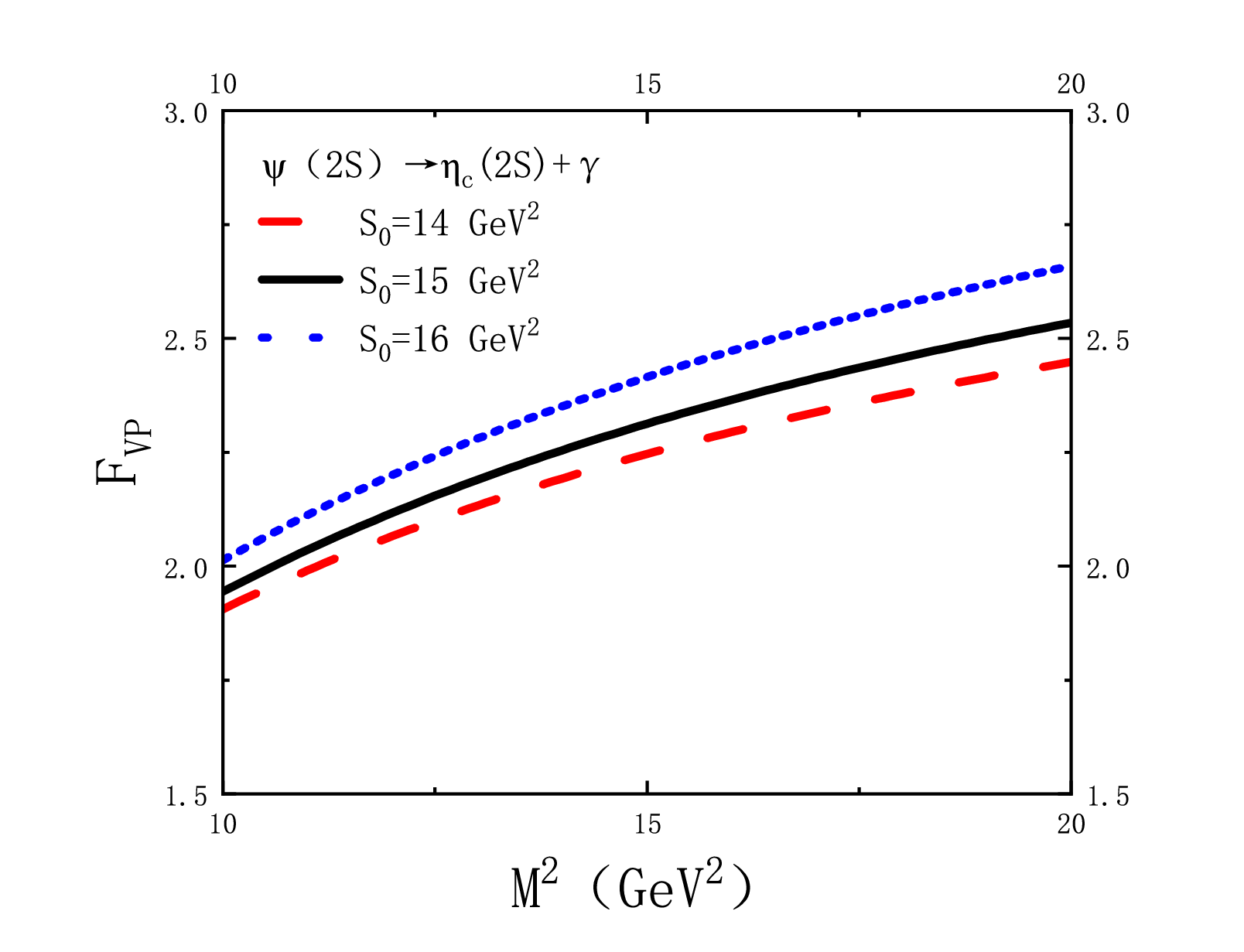}
        \label{Pf.eps}
    }
    \caption{The dependence of the form factors for (a) $K^{*-}\rightarrow K^-\gamma$ and (b) $\psi(2S)\rightarrow\eta_c(2S)\gamma$ processes on the threshold $s_0$ and Borel parameter $ M^2 $.}
    \label{dsdsdCCd}
\end{figure}

The form factor for the $K^{*-} \rightarrow K^- \gamma$ process is shown in Fig.~\ref{Kf.eps}, with the threshold parameter chosen as $s_0 \approx 1.9$ GeV$^2$. As illustrated in Fig.~\ref{Kf.eps}, our results exhibit high stability against variations in the threshold parameter and also remain highly stable under changes in the Borel parameter, indicating that our choice of the Borel region is reasonable.

Building on the transition form factors we have obtained for ground state vector mesons \cite{Sun:2010oyk}, we extend our calculation to excited state M1 radiative decay $\psi(2S) \rightarrow \eta_c(2S) \gamma$. As shown in Fig. \ref{Pf.eps}, the results of the form factor exhibits a certain sensitivity to the Borel parameter, while maintaining only weak dependence on the thresholds.

\begin{figure}[H]
    \centering
    \vspace{-0.50cm}
    \subfigure[]{
        \includegraphics[width=0.45\linewidth]{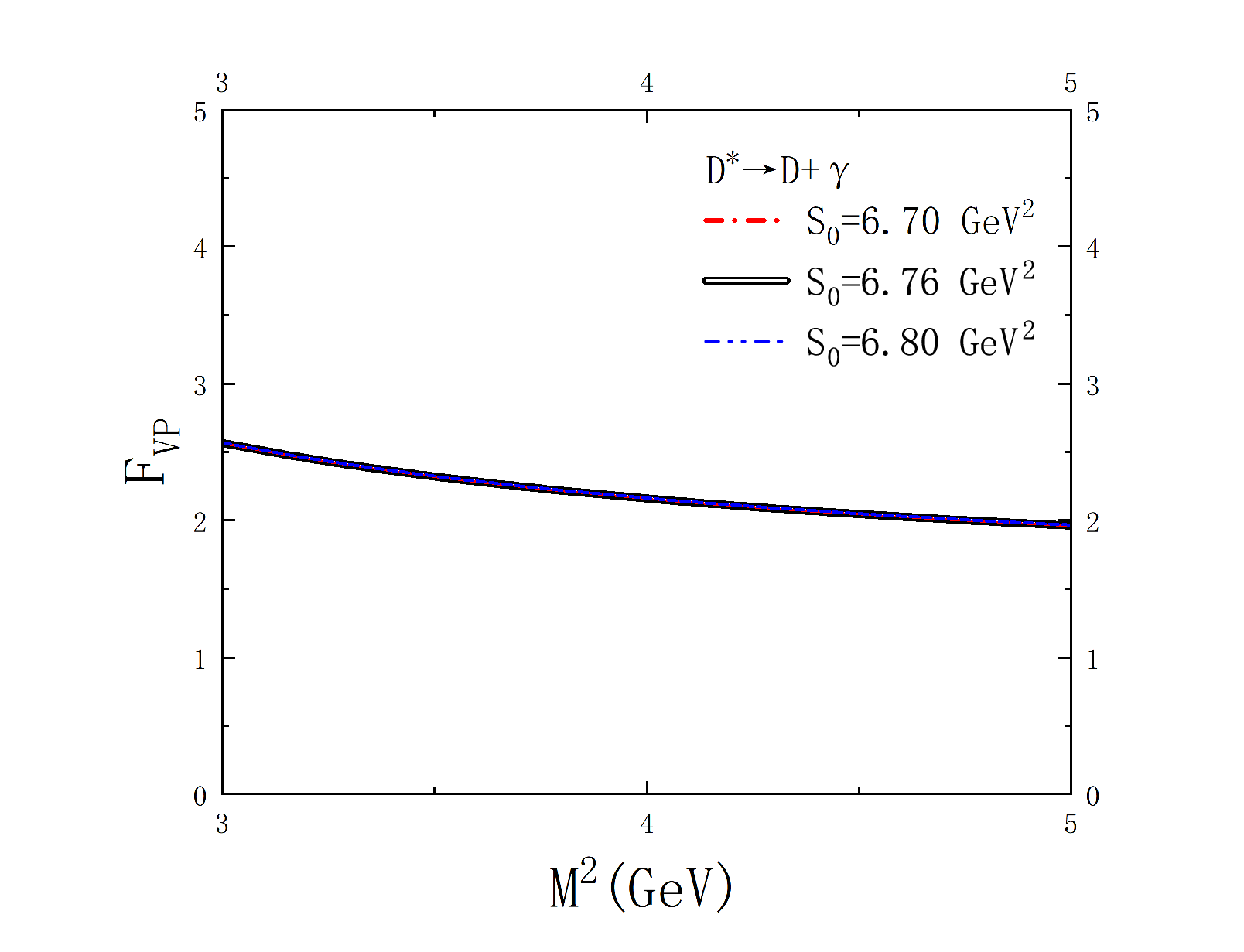}
        \label{Df.eps}
    }
    \quad
    \subfigure[]{
        \includegraphics[width=0.45\linewidth]{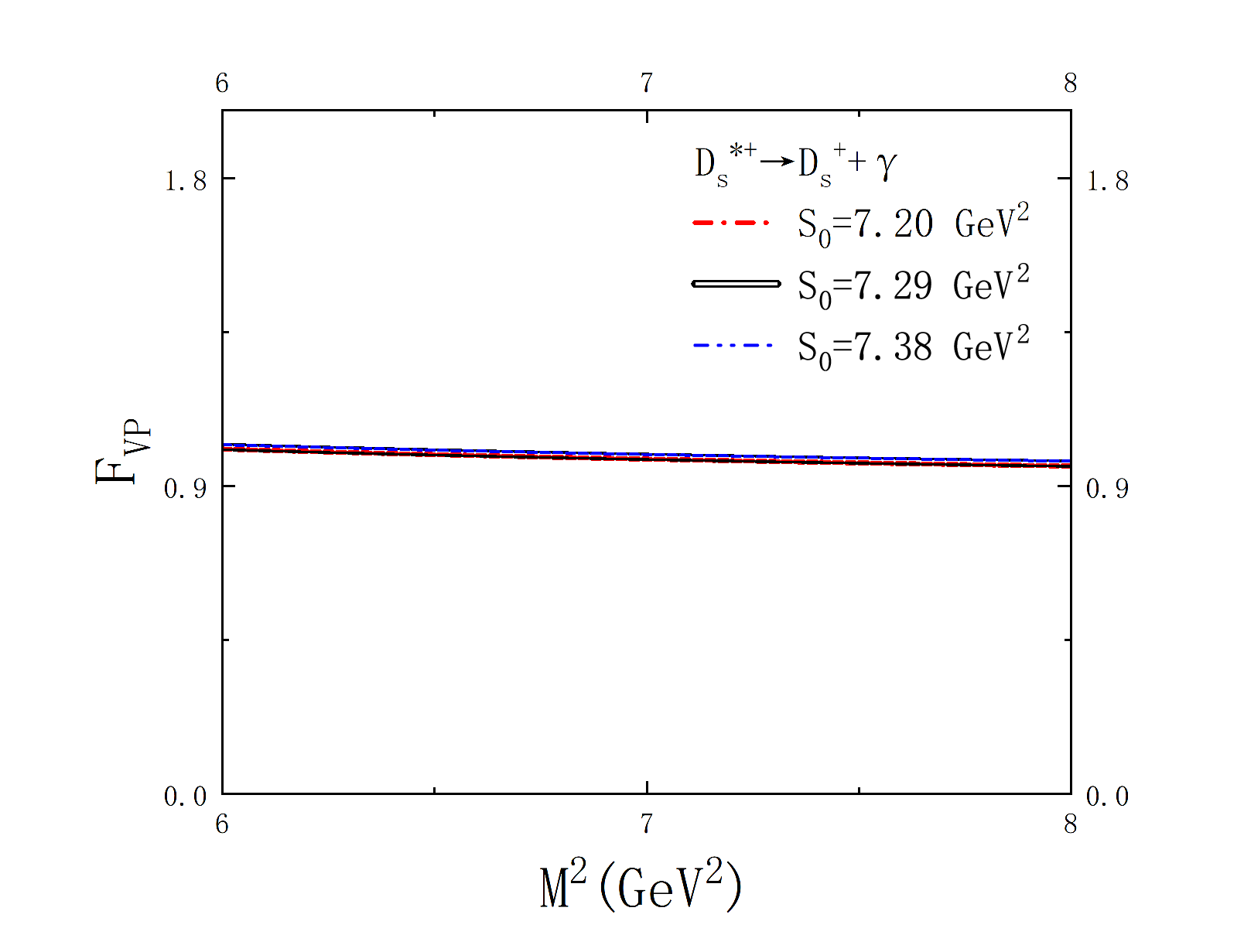}
        \label{DSf.eps}
    }
    \caption{The dependence of the form factors for (a) $D^*\rightarrow D\gamma$ and (b) $D^{*+}_s\rightarrow D^+_s\gamma$ processes on the threshold $s_0$ and Borel parameter $ M^2 $.}
    \label{dsdsdDDd}
\end{figure}

Experimental data on the $D^*\rightarrow D\gamma$ transition remain imprecise at present. In Fig. \ref{Df.eps}, we choose the threshold near the square of the mass of $D_1^{*0}(2600)$ \cite{ParticleDataGroup:2024cfk}. Within the LCSR framework, the form factor is remarkably stable against variations of the thresholds. By contrast, the results exhibit a certain dependence on the Borel parameter.

Similar to the $D^*\rightarrow D\gamma$ process, we select the threshold near the square of the $D^*_{s1}(2700)^{\pm}$ meson mass \cite{ParticleDataGroup:2024cfk}, $s_0 = 7.29$ GeV$^2$.
As shown in Fig. \ref{DSf.eps}, the transition form factor for $D_s^{*+}\to D_s^+\gamma$ exhibits excellent stability under variations of the thresholds, and the results of the form factor display only a mild dependence on the Borel parameter.

\begin{figure}[H]
    \centering
    \vspace{-0.50cm}
    \subfigure[]{
        \includegraphics[width=0.45\linewidth]{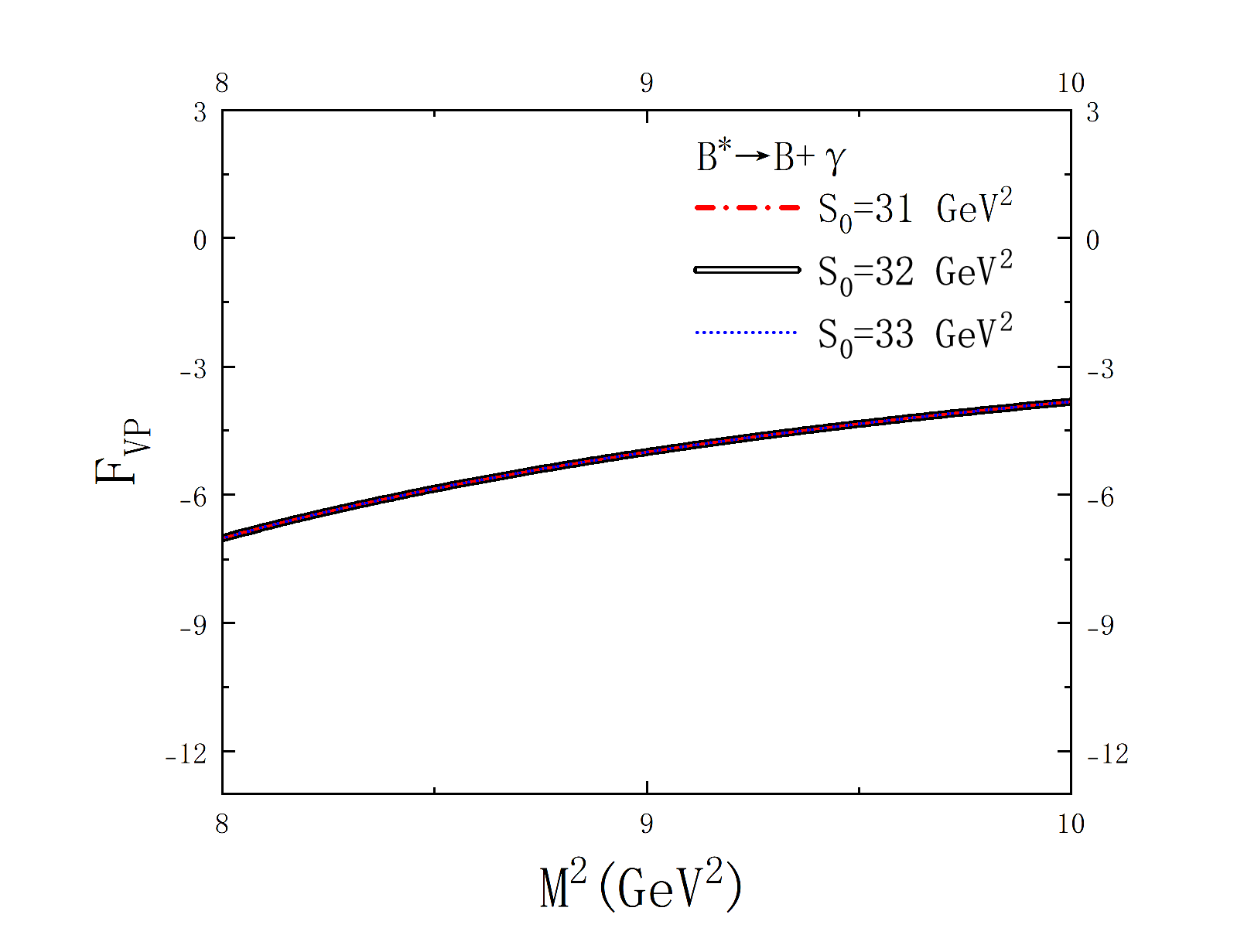}
        \label{Bf.eps}
    }
    \quad
    \subfigure[]{
        \includegraphics[width=0.45\linewidth]{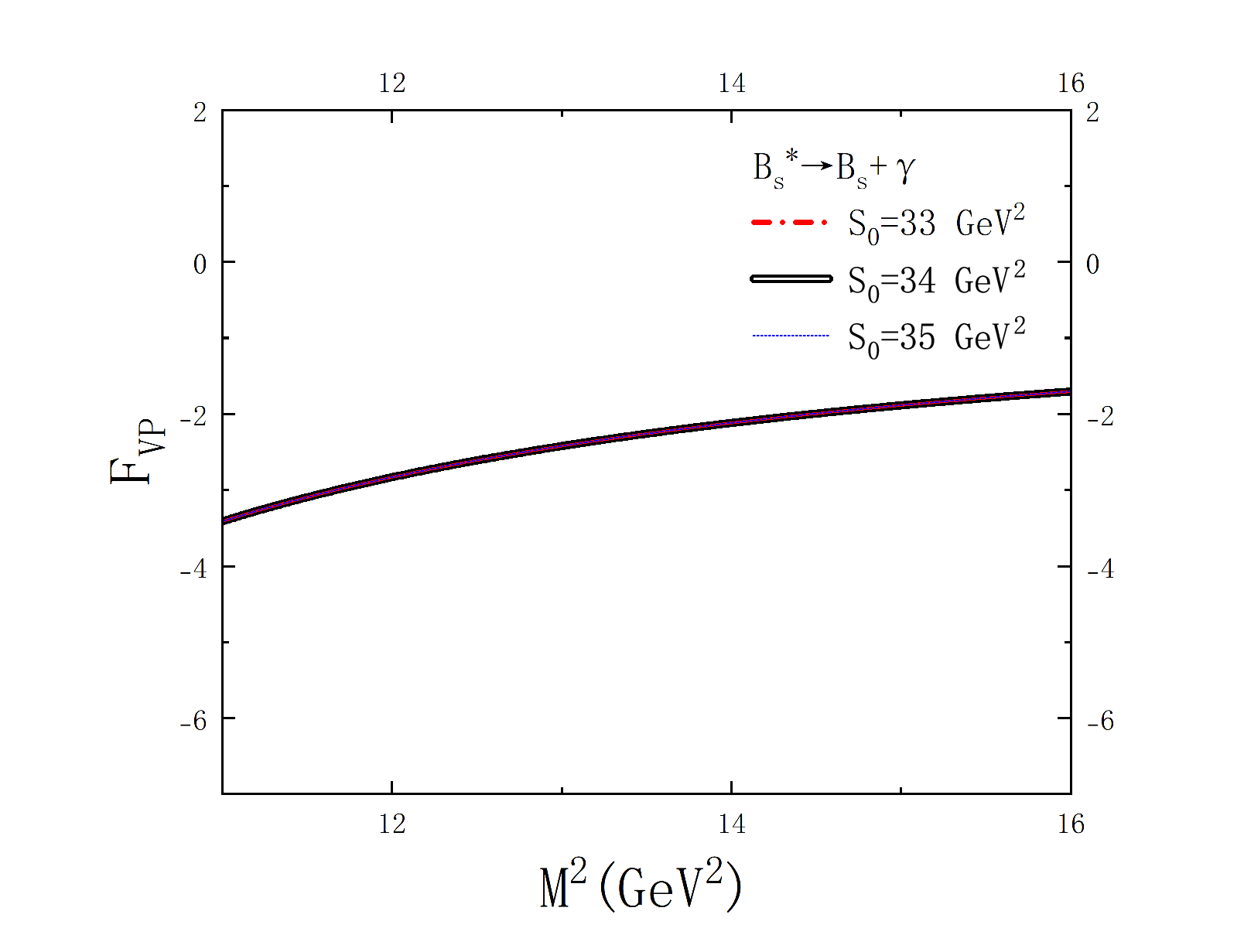}
        \label{BSf.eps}
    }
    \caption{The dependence of the form factors for (a) $B^*\rightarrow B\gamma$ and (b) $B^*_s\rightarrow B_s\gamma$ processes on the threshold $s_0$ and Borel parameter $ M^2 $.}
    \label{dsdsdpd}
\end{figure}

In experiments, the total decay widths of the $B^*$ and $B_s^*$ mesons and the branching fractions for their M1 radiative decays have not yet been precisely determined. As shown in Fig. \ref{dsdsdpd}, the transition form factors for the $B^*\rightarrow B\gamma$ and $B^*_s\rightarrow B_s\gamma$ processes exhibit a high degree of stability under variations of the Borel parameter, with almost negligible impact. Further, the form factors display weak dependence on the threshold. This minor variation stems primarily from the insufficient knowledge of the first excited state in these decay channels, which directly affects the choice of the threshold parameter and constitute the main source of uncertainty.

\subsection{Numerical results for the form factors $F_{VP}$}
\label{sec:bc}
Based on the above sensitivity analysis, we present the numerical values of the transition form factors at fixed Borel parameters and thresholds in Table \ref{tab:paEor} at the limit $q^2=0$. The total uncertainty incorporates contributions from various sources, including the Borel parameter $M^2$, the threshold parameter, quark masses, decay constants, and higher-twist distribution amplitudes. In the extraction of the form factors, the contributions from excited states are all below $20\%$, while those from higher-twist distribution amplitudes are less than $3\%$ and the radiative correction a few per mille. In this work, with the parameters choices $s_0 = 1.96$ GeV$^2$ and $M^2 = 5$ GeV$^2$, we present numerical results for the $K^{*-}\rightarrow K^-\gamma$  transition form factor. The parameters adopted for the other channels are  $D^*\rightarrow D\gamma$ ($s_0 = 6.76$ GeV$^2$, $M^2 = 4$ GeV$^2$), $B^*\rightarrow B\gamma$ ($s_0 = 32$ GeV$^2$, $M^2 = 9$ GeV$^2$), $D^{*+}_s\rightarrow D^+_s\gamma$ ($s_0 = 7.29$ GeV$^2$, $M^2 = 7$ GeV$^2$), $B_s^*\rightarrow B_s\gamma$ ($s_0 = 34$ GeV$^2$, $M^2 = 14$ GeV$^2$), and $\psi(2S)\rightarrow\eta_c(2S)\gamma$ ($s_0 = 16$ GeV$^2$, $M^2 = 15$ GeV$^2$).

\renewcommand{\tabcolsep}{0.3cm}
\renewcommand{\arraystretch}{1.0}
\begin{table}[h!]\centering
	\caption{Numerical results of the form factors $F_{VP}(q^2)$ at the limit $q^2\rightarrow 0$.}
	\label{tab:paEor}
	\begin{tabular}{ccccccc}
		\bottomrule[1.0pt]\bottomrule[0.5pt]
		& $K^{*-}\rightarrow K^-\gamma$ & $D^*\rightarrow D\gamma$ & $D^{*+}_s\rightarrow D^+_s\gamma$ & $B^*\rightarrow B\gamma$  & $B_s^*\rightarrow B_s\gamma$ & $\psi(2S)\rightarrow\eta_c(2S)\gamma$ \\\hline
		$F_{VP}(q^2=0)$ & $0.569\pm0.106$ & $2.16\pm0.41$ & $0.986\pm0.030$  & $-4.996\pm2.100$ & $-2.26\pm1.16$ & $2.41\pm0.06$ \\
		\bottomrule[0.5pt]\bottomrule[1.0pt]
	\end{tabular}
\end{table}

In this work, we define $q^2$ as the four-momentum squared of the photon, i.e., the photon virtuality. For the radiative decays $V\rightarrow P+ \gamma$ considered here, the photon is real, which corresponds to the physical point $q^2=0$. Our light-cone sum rule calculations directly yield the form factors at this point, denoted as $F_{VP}(0)$ and listed in Table \ref{tab:paEor}. These values can be directly inserted into the decay width formula Eq. (\ref{for:OBQQQ}) , enabling a direct comparison with experimental measurements.

Concerning the $q^2$ dependence, we note that the form factor  $F_{VP}(q^2)$ as a function of photon virtuality is primarily relevant for processes involving virtual photons, such as $V \rightarrow P+l^{+}+l^{-}$. In such cases, a single-pole parametrization motivated by the vector meson dominance (VMD) model is commonly adopted to describe this dependence. For the real photon decay $V\rightarrow P+ \gamma$, however, the physical kinematic point is $q^2=0$, which is consistent with $0 \leq q^2\leq (M_V-M_P)^2$ in the LCSR, and therefore no extrapolation is required.

\subsection{Numerical results for the decay widths}
\label{sec:bQc}

In the aforementioned section, we discussed the transition form factors, an important nonperturbative physical parameter. In this section, we will present the results of the M1 radiative decay widths using the transition form factors. We visually illustrate the dependence of M1 radiative decay widths on the thresholds $s_0$ and Borel parameters $M^2$.

\begin{figure}[H]
    \centering
    \vspace{-0.50cm}
    \subfigure[]{
        \includegraphics[width=0.45\linewidth]{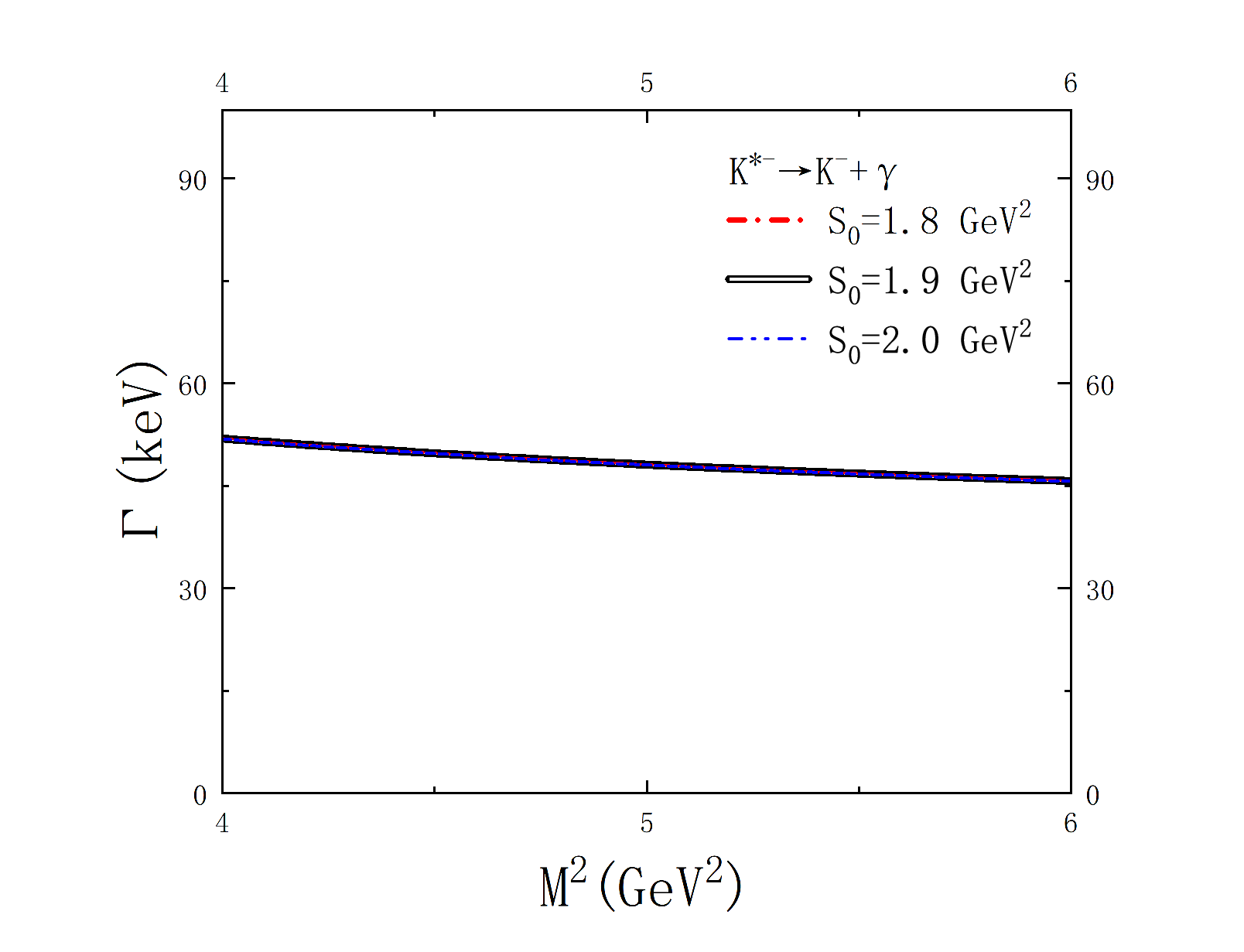}
        \label{Kk.eps}
    }
    \quad
    \subfigure[]{
        \includegraphics[width=0.45\linewidth]{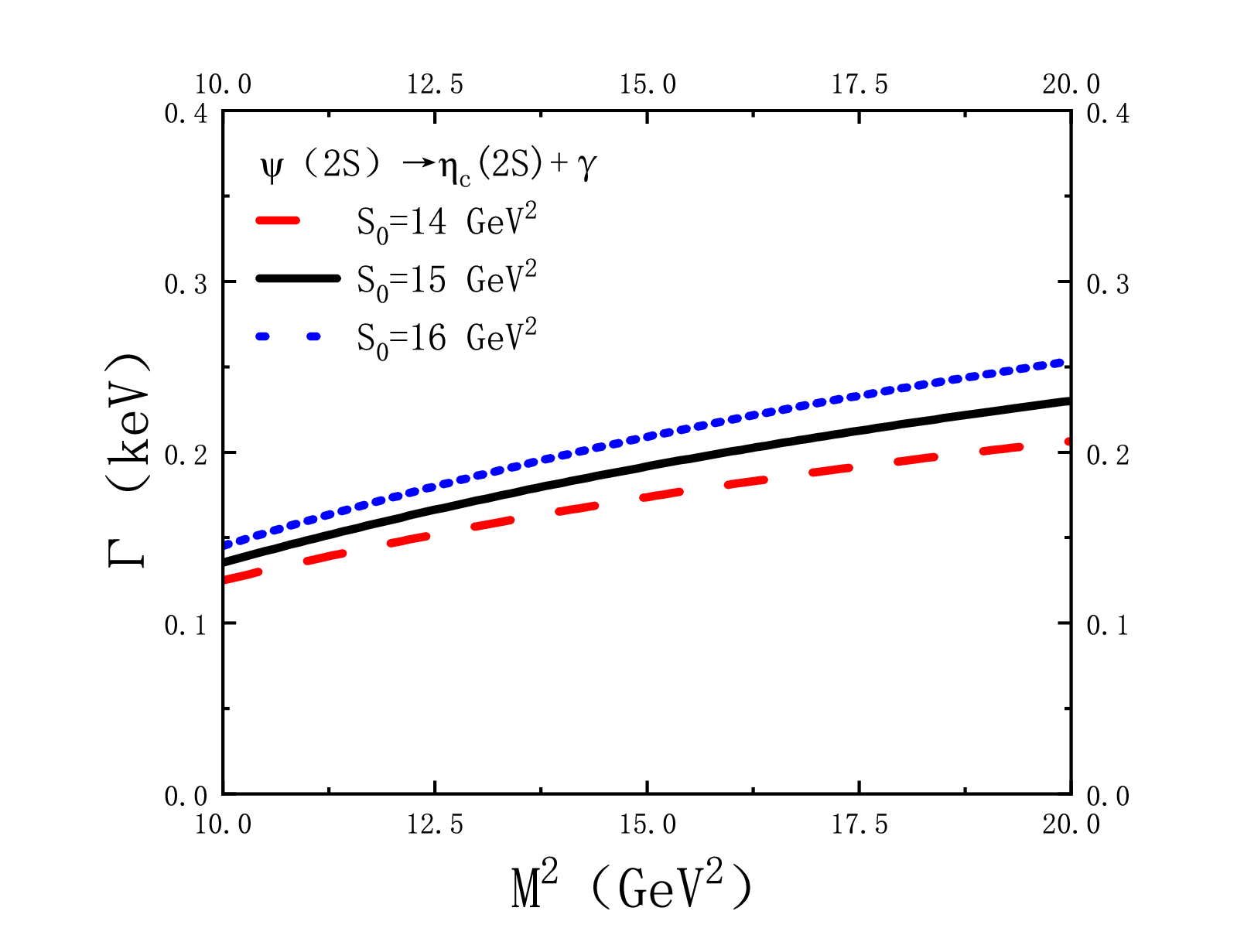}
        \label{Pk.eps}
    }
    \caption{The dependence of the decay widths for (a) $K^{*-}\rightarrow K^-\gamma$ and (b) $\psi(2S)\rightarrow\eta_c(2S)\gamma$ processes on the threshold $s_0$ and Borel parameter $ M^2 $.}
    \label{Tdsdsdd}
\end{figure}

The Fig. \ref{Tdsdsdd} shows the dependence of the decay widths for the processes $K^{*-}\to K^{-}\gamma$ and $\psi(2S)\to \eta_c(2S)\gamma$ on the thresholds $s_0$ and the Borel parameters $M^2$. It is evident from the figure that the M1 radiative decay width remains remarkably stable under variations of both the threshold parameter and the Borel parameter. The Appendix presents plots illustrating the continuous variation of selected decay widths with the thresholds $s_0$ and the Borel parameters $M^2$; further details are provided in  the Appendix.

\begin{figure}[H]
    \centering
    \vspace{-0.50cm}
    \subfigure[]{
        \includegraphics[width=0.45\linewidth]{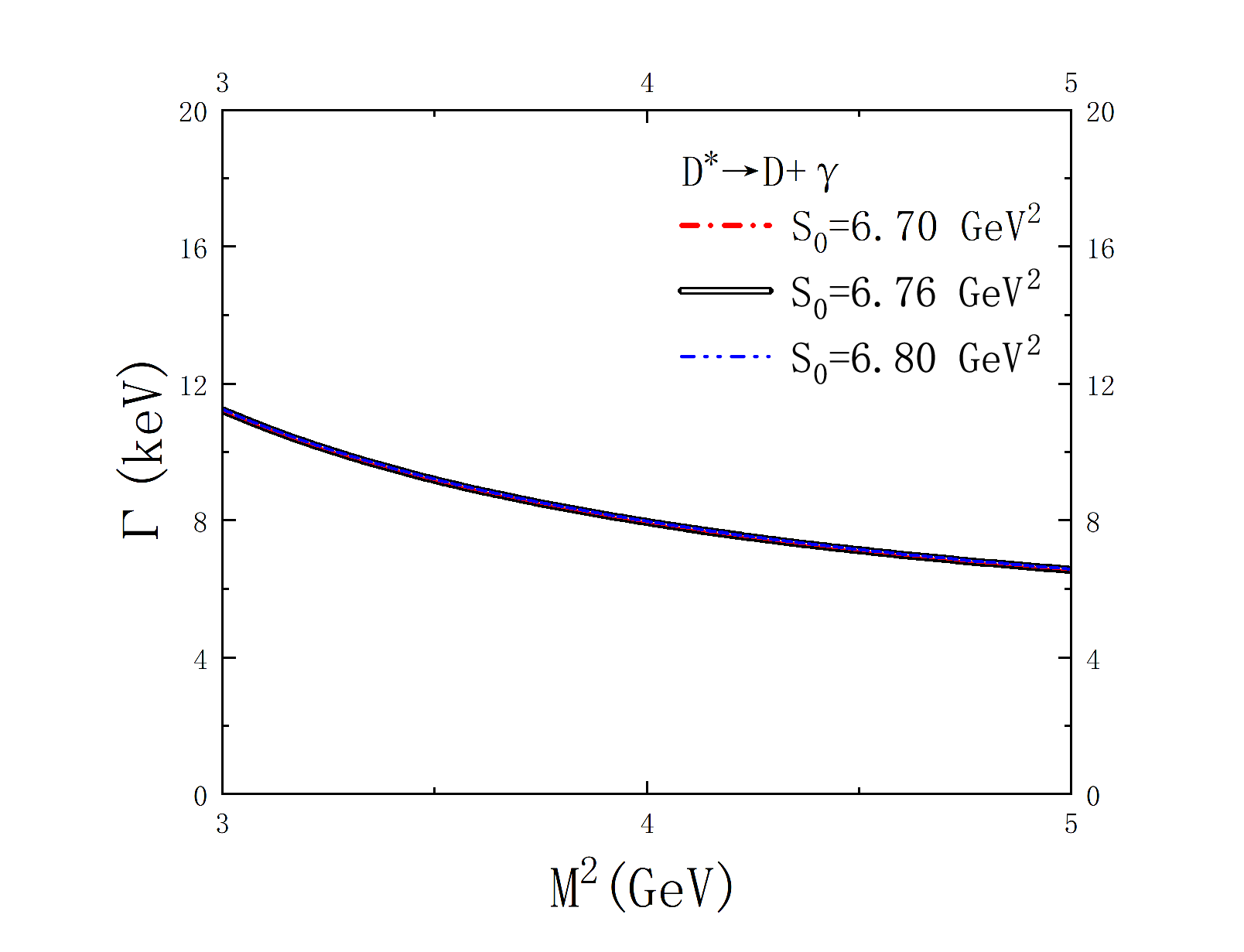}
        \label{Dk.eps}
    }
    \quad
    \subfigure[]{
        \includegraphics[width=0.45\linewidth]{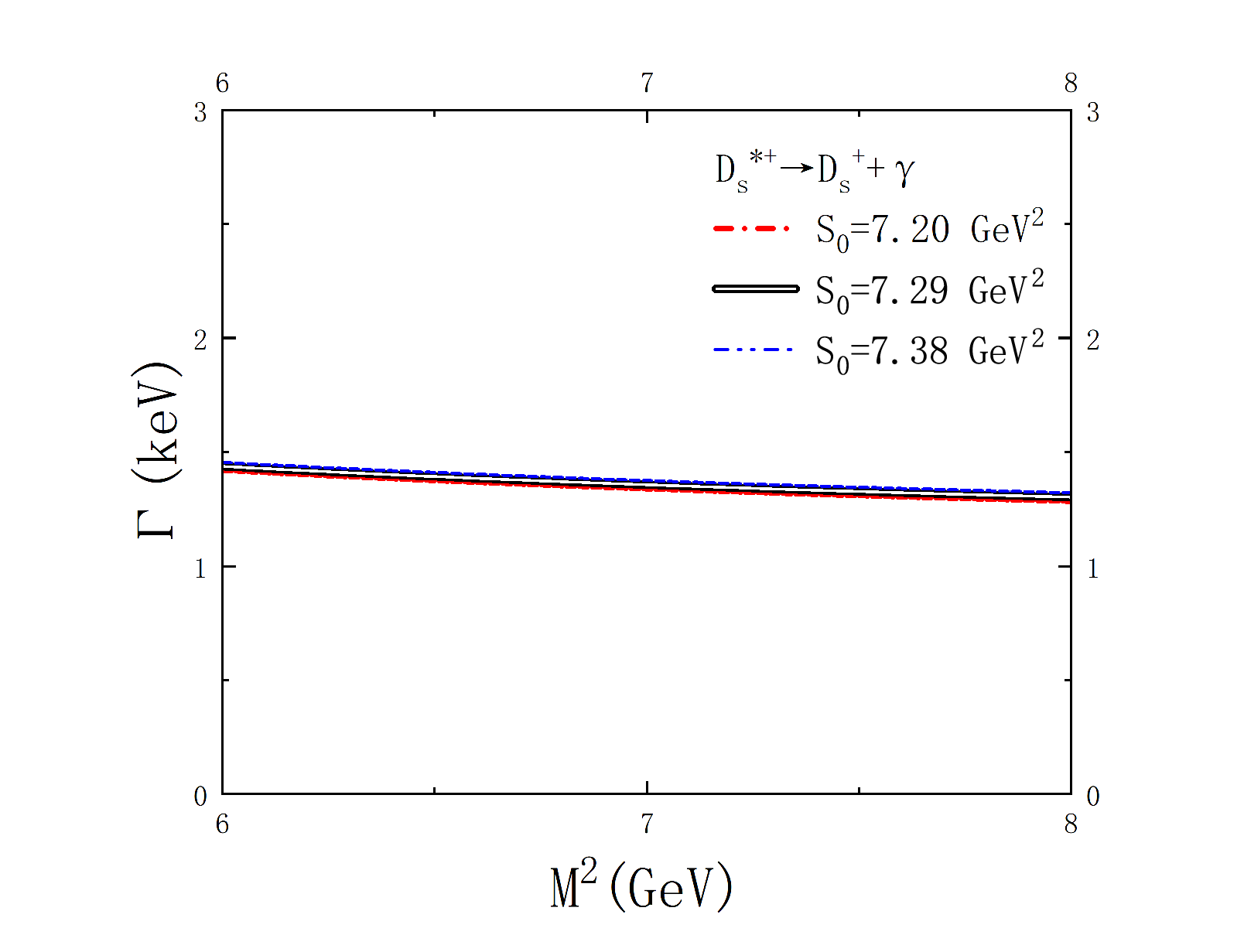}
        \label{DSk.eps}
    }
    \caption{The dependence of the decay widths for (a) $D^*\rightarrow D\gamma$ and (b) $D^{*+}_s\rightarrow D^+_s\gamma$ processes on the threshold $s_0$ and Borel parameter $ M^2 $.}
    \label{dsdsddD}
\end{figure}

\begin{figure}[H]
    \centering
    \vspace{-0.50cm}
    \subfigure[]{
        \includegraphics[width=0.45\linewidth]{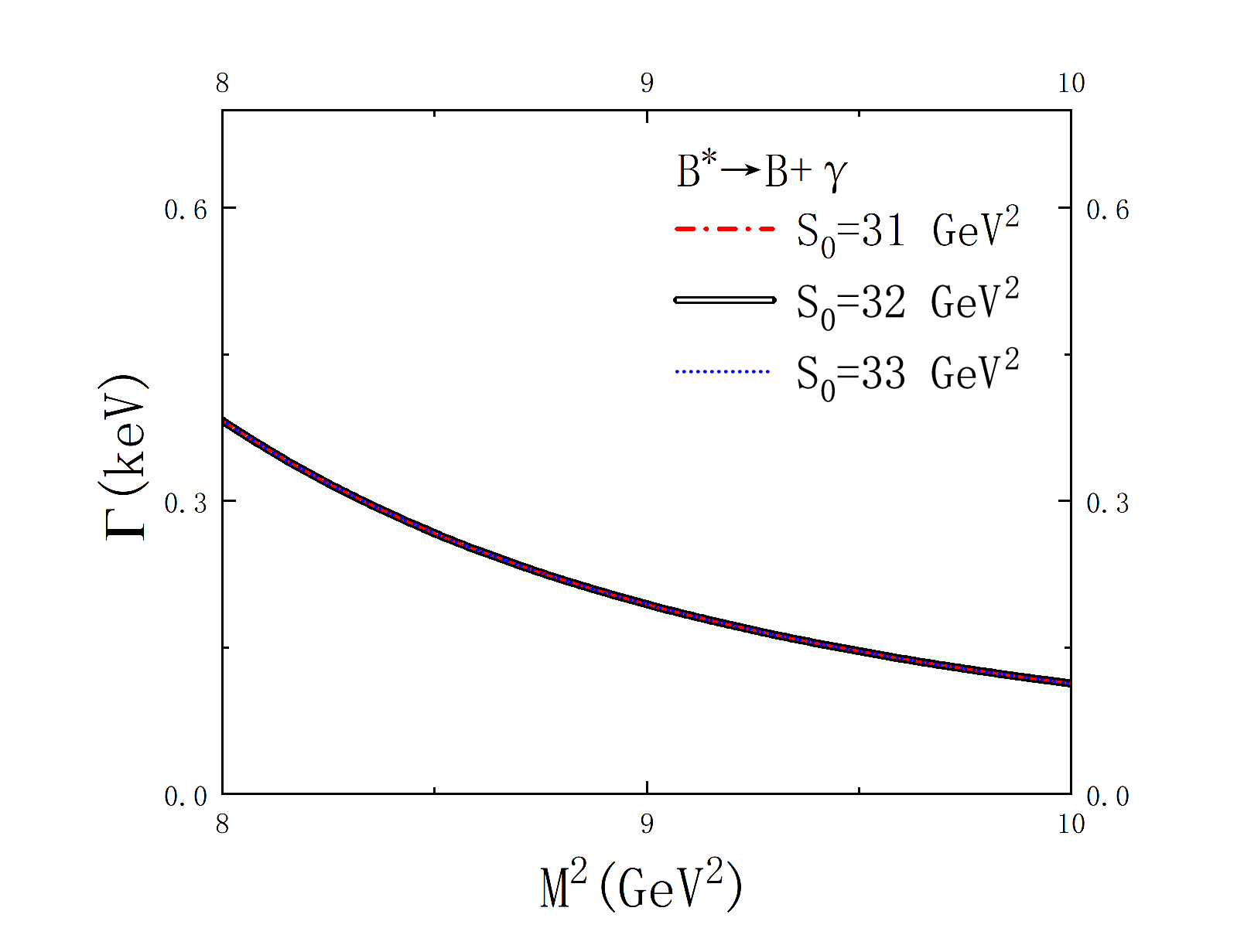}
        \label{Bk.eps}
    }
    \quad
    \subfigure[]{
        \includegraphics[width=0.45\linewidth]{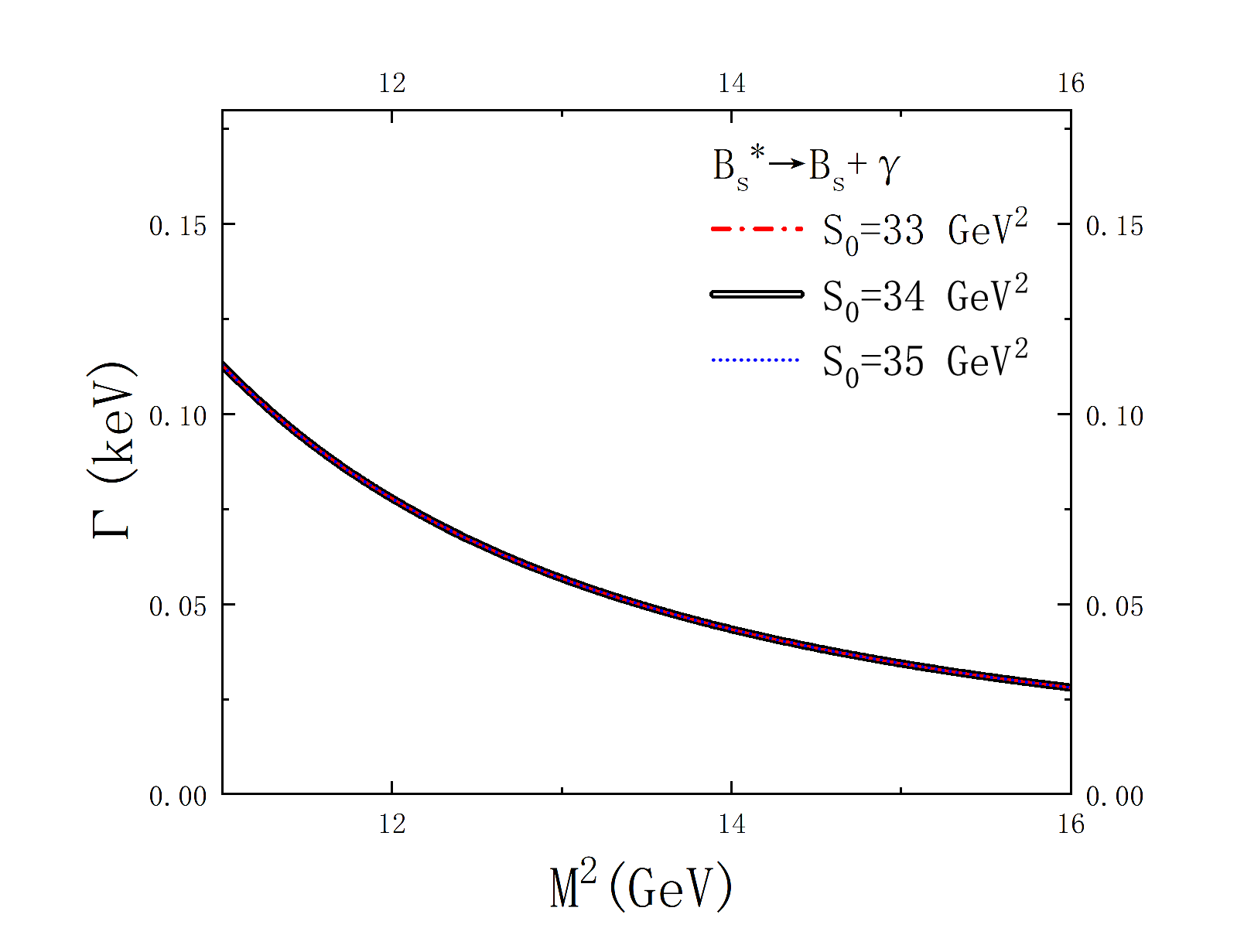}
        \label{BSk.eps}
    }
    \caption{The dependence of the decay widths for (a) $B^*\rightarrow B\gamma$ and (b) $B_s^*\rightarrow B_s\gamma$ processes on the threshold $s_0$ and Borel parameter $ M^2 $.}
    \label{dsdsddR}
\end{figure}

The M1 radiative decay width for the $K^{*-} \rightarrow K^-\gamma$ process, as quoted in the Particle Data Group (PDG) review, is $50.37$ keV \cite{ParticleDataGroup:2024cfk}. The result obtained in this work using the LCSR method is $48.10\pm15.05$ keV, with a relative deviation of approximately $4\%$ compared to the PDG value. For the excited state $\psi(2S) \rightarrow \eta_c(2S)\gamma$ process, the calculated result in this work is $0.20\pm0.07$ keV, and the relative deviation from the experimental value of $0.15$ keV \cite{BESIII:2023lcc} is approximately $33\%$. This discrepancy could be attributed to several factors, including the neglect of higher-order corrections, the sensitivity of the excited-state calculation to the LCSR parameters and the challenges in modeling the LCDA for the $\eta_c(2S)$ meson.

Theoretical predictions for the M1 radiative decay widths of four processes ($D^*\rightarrow D\gamma$, $B^*\rightarrow B\gamma$, $D^{*+}_s\rightarrow D^+_s\gamma$ and $B_s^*\rightarrow B_s\gamma$) are presented in Figs. \ref{dsdsddD} and \ref{dsdsddR}. In Table \ref{tab:parsa}, we compare our results of the decay widths with those obtained from experiments and from  other theoretical approaches,  with the errors in our results coming from the Borel parameter, the threshold parameter, quark masses, decay constants, and higher-twist distribution amplitudes.

\renewcommand{\tabcolsep}{0.35cm}
\renewcommand{\arraystretch}{1.0}
\begin{table}[H]
	\caption{The M1 radiative decay widths for all processes are presented (in units of keV). The first row shows our theoretical predictions, and the second row displays the corresponding experimental measurements. Furthermore, we systematically compare our theoretical results with predictions from several models, including QCD sum rules (QCDSR), vector confining potentials (VCP), relativistic quark model(RQM), nonrelativistic potential model(PM), effective
 mass scheme(EMS), scalar confining potentials (SCP), light-front quark model (LFQM), and nonrelativistic quark model (NRQM).}
 \label{tab:parsa}
 \begin{tabular}{lcccccc}
 	\bottomrule[1.0pt]\bottomrule[0.5pt]
 	& $K^{*-}\rightarrow K^-\gamma$ & $D^*\rightarrow D\gamma$ & $D^{*+}_s\rightarrow D^+_s\gamma$ & $B^*\rightarrow B\gamma$  & $B_s^*\rightarrow B_s\gamma$ & $\psi(2S)\rightarrow\eta_c(2S)\gamma$ \\\hline
 	
 	This work       & $48.10\pm15.05$ & $7.96\pm3.31$ & $1.36\pm0.08$ & $0.194\pm0.200$  & $0.05\pm0.06$ & $0.20\pm0.07$ \\
 	$EXP$ \cite{ParticleDataGroup:2024cfk,BESIII:2023lcc}  &50.37 & ... & ... & ... & ... & $0.15$\\
 	$VCP$ \cite{Ebert:2002xz}   & ... &14.3 &... &0.087  &0.064 &...\\
 	$SCP$ \cite{Ebert:2002xz}   &... &17.4 &... &0.101  &0.074 &...  \\
 	$RIQ$ \cite{Priyadarsini:2016tiu}   & ... &26.5 &0.213 &0.181  &0.119 &...\\
 	$LFQM$ \cite{Choi:2007se,Peng:2012tr,Choi:1997iq}  &79.5 &20.0$\pm$0.3 &0.17 &0.13$\pm$ 0.01  &0.068$\pm$ 0.017 &$0.11$\\
 	$NRQM$ \cite{Ebert:2002xz,Li:2011ssa} & ... &37 &... &0.27  &0.132 &$0.21$\\
 	$LCSR$ \cite{Aliev:1995zlh}   & ... &14.4 &... &0.16	 &... &...\\
 	$QCDSR$ \cite{Zhu:1996qy} & ... &12.9$\pm$2 &... &0.13$\pm0.03$  &$0.22$$\pm$$0.04$ &...\\
 	$HQET$ \cite{Colangelo:1993zq}&... &16.0$\pm$7.5 &... &0.075$\pm$0.027  &... & \\
    $PM$ \cite{mohan2025systematicstudylightcharm,Bonnaz:2001aj} &104.46 &35.506 &0.284 &... &... &... \\
    $EMS$ \cite{mohan2025systematicstudylightcharm} &62.988 &11.731 &0.130 &... &... &... \\
    $RQM$ \cite{Li:2022vby} &... &97 &... &... &... &... \\
 	\bottomrule[0.5pt]\bottomrule[1.0pt]
 \end{tabular}
\end{table}

We observe that, as we move from $K^{*}$ to the $ D^*, D_s^{*}, \psi(2S), B^*$ and  $B_s^{*}$ mesons, the meson masses increase successively, while the decay widths exhibit an overall decreasing trend. This behavior reflects the mass suppression effect characteristic of heavy quarks, which is physically expected and consistent with the predictions of heavy quark effective theory \cite{Zhang:2025wmr, Liu:2025fbe}. Furthermore, this trend is in good agreement with the patterns shown by other theoretical approaches, such as $RIQ$ \cite{Priyadarsini:2016tiu}, $LFQM$ \cite{Choi:2007se,Peng:2012tr,Choi:1997iq}, $PM$ \cite{mohan2025systematicstudylightcharm,Bonnaz:2001aj}, and $EMS$ \cite{mohan2025systematicstudylightcharm}, as presented in Table \ref{tab:parsa}. The radiative decay widths for $D^*$ and $B^*$ mesons predicted by our LCSR approach are in good agreement with those from other LCSR calculations within uncertainties. For the charmonium excited state, we present the first LCSR calculation and find that our result is consistent within uncertainty with the prediction from the nonrelativistic quark model.

\subsection{Linear relationship}
\label{sec:bcV}
An intriguing systematic behavior emerges from our analysis of the ground-state $V\rightarrow P \gamma$ decays. We find that the decay widths exhibit a universal scaling relation when plotted against a specific kinematic function. Motivated by the two-body decay dynamics and the role of decay constant, we define the function A(x) as

\begin{equation}\label{for:fit}
	\begin{aligned}
	A(x)=\left( \frac{f_{f}}{f_{i}} \frac{1}{(m_{f}+m_{i})^2} \right) k_\gamma^3 (F_{VP})^x
	\end{aligned}
\end{equation}
where, $x$ is a free power to be determined, $f_{i}$ and $f_{f}$ are the decay constants of the initial and final states, $m_{i}$ and $m_{f}$ are the masses corresponding to the initial and final state mesons and $k_{\gamma}$ is the photon momentum.

\begin{figure}[H]
    \centering
    \vspace{-0.50cm}
    \subfigure{
        \includegraphics[width=1.00\linewidth]{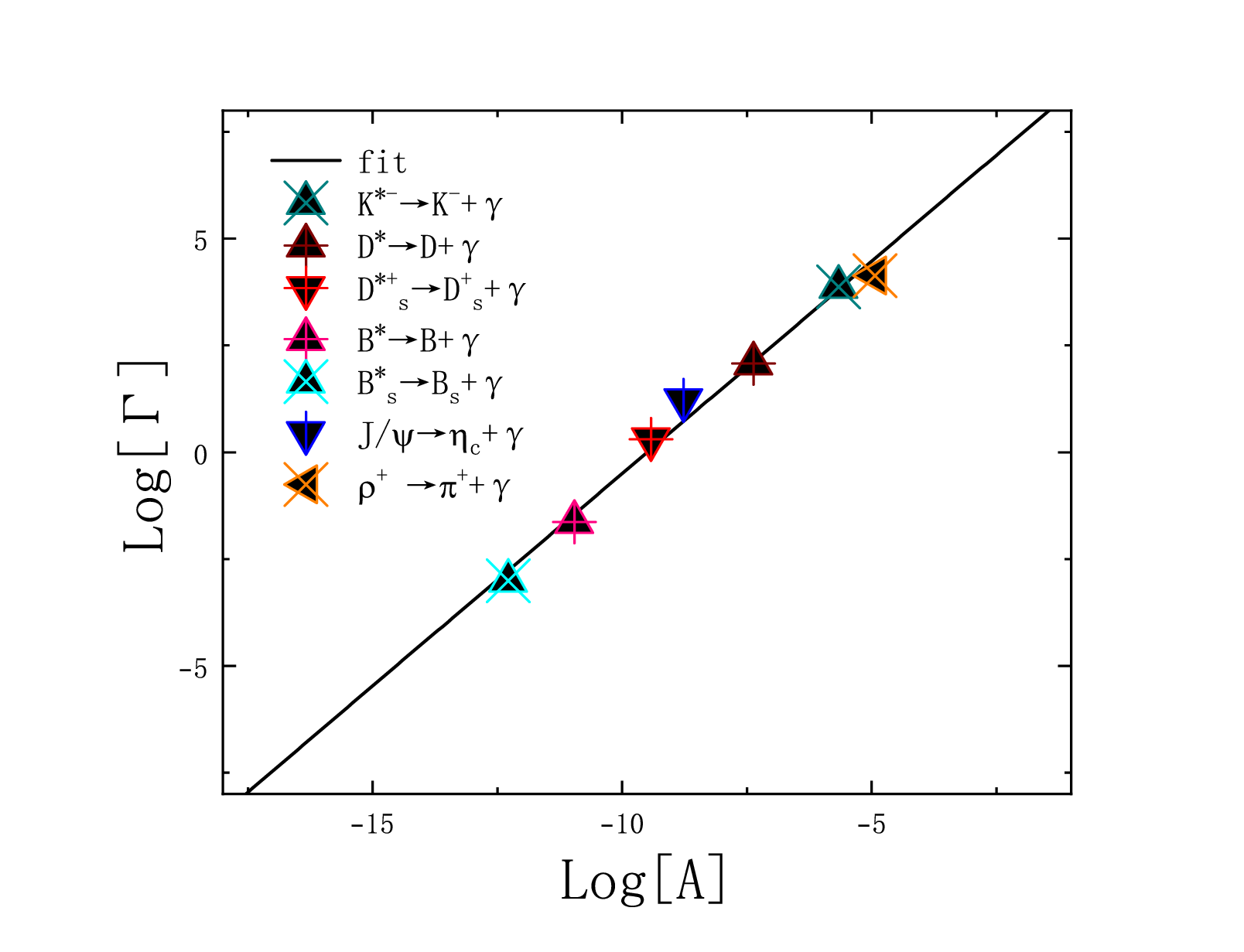}
        \label{Fit7.eps}
    }
    \caption{Linear dependence of the decay width on the function A in the logarithmic coordinate system.}
    \label{Fit7.eps}
\end{figure}

Remarkably, when plotting the decay width $\Gamma$ against $A(x)$ in the logarithmic coordinate system, all data points for the ground-state transitions align along a straight line for an optimal power of $x\approx2.1$, as shown in Fig. \ref{Fit7.eps}. This observed linear dependence, log($\Gamma$)  $\propto$  log($A(2.1)$) suggests a universal pattern governing these M1 radiative decays. The fact that the optimal power is not exactly 2[as a naive leading-order analysis might suggest in Eq. (\ref{for:OBQQQ})] indicates the significance of contributions beyond the leading order, such as higher-order perturbative QCD corrections and higher-twist effects. Additionally, we have also collected the meson masses, decay constants, radiative decay widths, and transition form factors for the processes $J/\psi \rightarrow \eta_c\gamma$ and $ \rho^+ \rightarrow \pi^+ \gamma $  \cite{Zhang:2015mxa,Guo:2019xqa,ParticleDataGroup:2024cfk}, also fall neatly on the same fitted line. This observation could provide a fresh, unifying perspective on the systematics of radiative decays.

\section{SUMMARY}
\label{sec:summary}

In this work, we study the magnetic dipole (M1) radiative decays of vector mesons within the framework of LCSR. Firmly grounded in the first principles of QCD, this method provides a computationally efficient tool for determining transition form factors with minimal model dependence. We apply it to a set of vector mesons systems spanning the strange, charm, and bottom sectors ($K^*$, $B^*$, $B_s^*$, $D^*$, $D^{*+}_s$) and extend the formalism to the radiative transition between excited charmonium states, $\psi(2S)\rightarrow\eta_c(2S)\gamma$. The core of this work lies in the computation of the transition form factors for these decay channels, from which the corresponding M1 radiative decay widths are derived.

Our numerical results of the decay widths are consistent with existing experimental data. For the ground state vector meson channel $K^{*-}\rightarrow K^-\gamma$, we predict $\Gamma(K^{*-}\rightarrow K^-\gamma)=48.10\pm15.05$ keV, in excellent agreement with the measured value of $50.37$ keV \cite{ParticleDataGroup:2024cfk}. For the charmonium transition $\psi(2S)\rightarrow\eta_c(2S)\gamma$, we obtain $\Gamma(\psi(2S)\rightarrow\eta_c(2S)\gamma)=0.20\pm0.07$ keV, which is consistent with the experimental value of $0.15$ keV \cite{BESIII:2023lcc} within uncertainties. For channels not yet measured experimentally, we provide the following theoretical predictions: $\Gamma(D^*\rightarrow D\gamma)=7.96\pm3.31$ keV, $\Gamma(D^{*+}_s\rightarrow D^+_s\gamma)=1.36\pm0.08$ keV, $\Gamma(B^*\rightarrow B\gamma)=0.194\pm0.200$  keV, and $\Gamma(B^*_s\rightarrow B_s\gamma)=0.05\pm0.06$ keV. A comprehensive comparison with a range of other theoretical approaches including quark models, lattice QCD, and alternative QCD sum rule calculations shows broad consistency, thereby further reinforcing the credibility of our predictions.

While this work provides individual predictions for decay widths, it also reveals a linear behavior governing these transition processes. For the vector mesons ($K^*$, $D^*$, $D^{*+}_s$, $\psi(2S)$, $B^*$, $B_s^*$), the decay width exhibits a monotonic decrease with increasing vector meson mass. This is consistent with the expectations of methods such as heavy quark effective theory. Moreover, when analyzed in logarithmic coordinates, the widths for all channels including charmonium systems reveal a striking linear scaling relation with an appropriate kinematic function derived from two-body decay dynamics. This finding suggests a potential universality underlying M1 transitions. Deviations from the fitted linear relation are primarily attributed to the leading order truncation in our computational framework, particularly the omission of higher order perturbative QCD corrections and higher twist contributions.

However, the physical origin of this scaling behavior remains to be clarified. Whether it reflects an underlying symmetry or common dynamical features across different flavor sectors requires further investigation. To elucidate its nature and test its generality, extending our analysis to other radiative transitions, such as electric dipole decays, is expected in the future. These extended studies will help determine whether the observed linear scaling is specific to M1 transitions or a more general manifestation of nonperturbative QCD dynamics.

In summary, the light-cone sum rules framework has been employed to compute radiative partial widths for a representative set of vector meson decays, yielding results consistent with experimental data while providing valuable physical insights. The observed linear behavior offers a new perspective on the systematics of hadronic radiative transitions. In subsequent work, we will focus on incorporating higher order corrections and related factors to enhance the precision of our results and reduce theoretical uncertainties. Furthermore, we plan to include processes with different spin configurations, thereby enabling more rigorous tests of the linear behavior observed in the present study. On the experimental front, high precision measurements of radiative widths for bottom and charmed mesons from ongoing and future facilities, particularly Belle II and LHCb, are anticipated to provide critical tests of these theoretical predictions and further constrain nonperturbative QCD models. Such coordinated efforts will undoubtedly advance our fundamental understanding of strong interaction dynamics in the nonperturbative regime.

\medskip
\section*{ACKNOWLEDGMENTS}

We would like to thank Wen-Nian Liu, Chen Wang, and Wen-Xuan Zhang for stimulating discussions. D. G. acknowledges the support of the China Scholarship Council program (Project ID: 202510530001).

\medskip
\section*{DATA AVAILABILITY}
The data that support the findings of this article are not publicly available. The data are available from the authors upon reasonable request.
\appendix
\section*{Appendix}
\label{sec:appendix}
In Sec. \ref{sec:fTToor}, we present a detailed discussion of the choice of the Borel parameter $M^2$ and the continuum threshold $s_0$, and we provide a quantitative assessment of how the transition form factors and the M1 radiative decay widths depend on these inputs. The figures in Sec. \ref{sec:fTToor}  illustrate the impact of continuous variation of the Borel parameter. For the threshold parameter, we analyze several representative values. For completeness and generality, the Appendix  provides a detailed account of the dependence of the transition form factors and the decay widths on the parameters $M^2$ and $s_0$ for all processes considered in this work.

For the process $K^*\rightarrow K^+\gamma$,
\begin{figure}[H]
    \centering
    \vspace{-0.50cm}
    \subfigure[]{
        \includegraphics[width=0.45\linewidth]{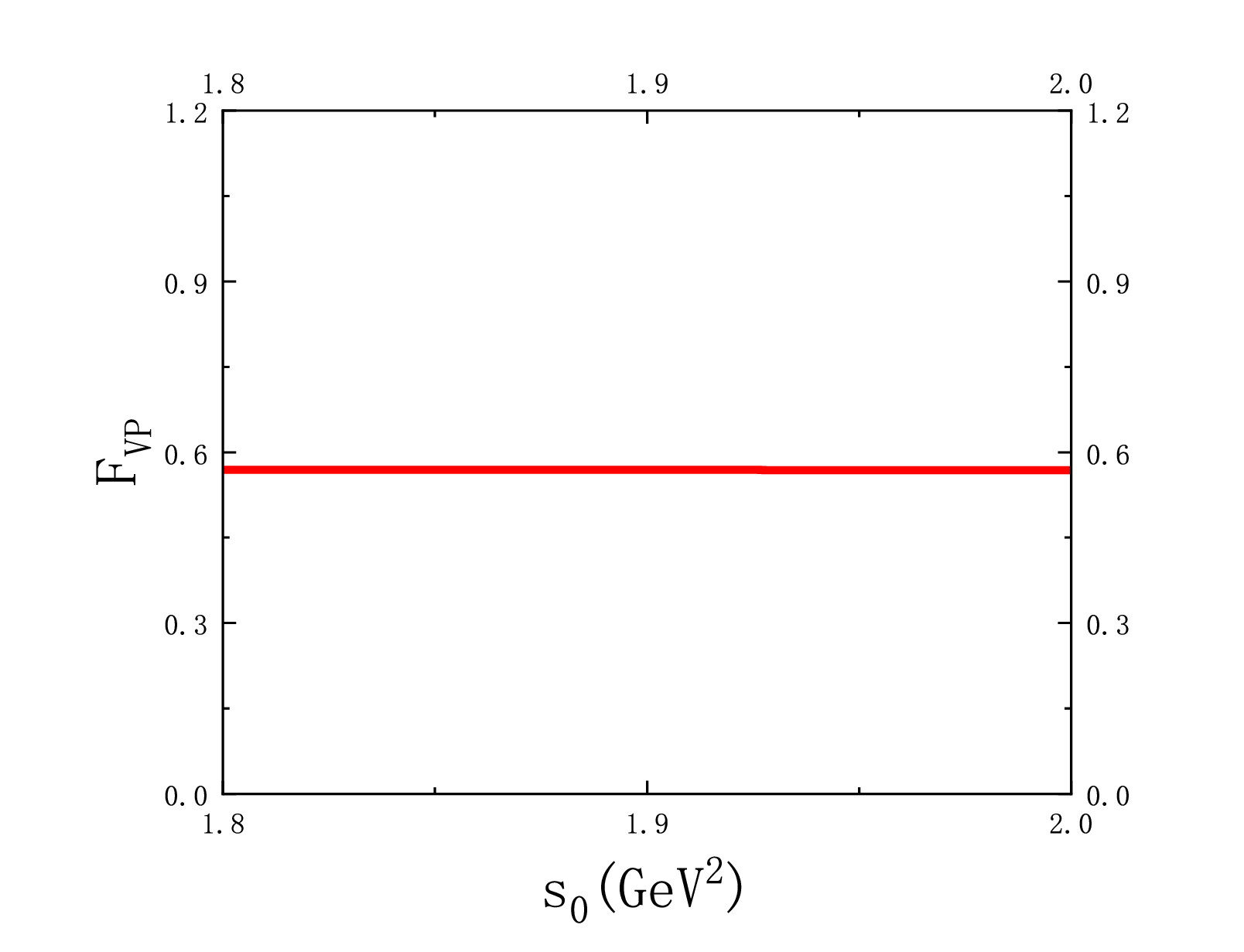}

    }
    \quad
    \subfigure[]{
        \includegraphics[width=0.45\linewidth]{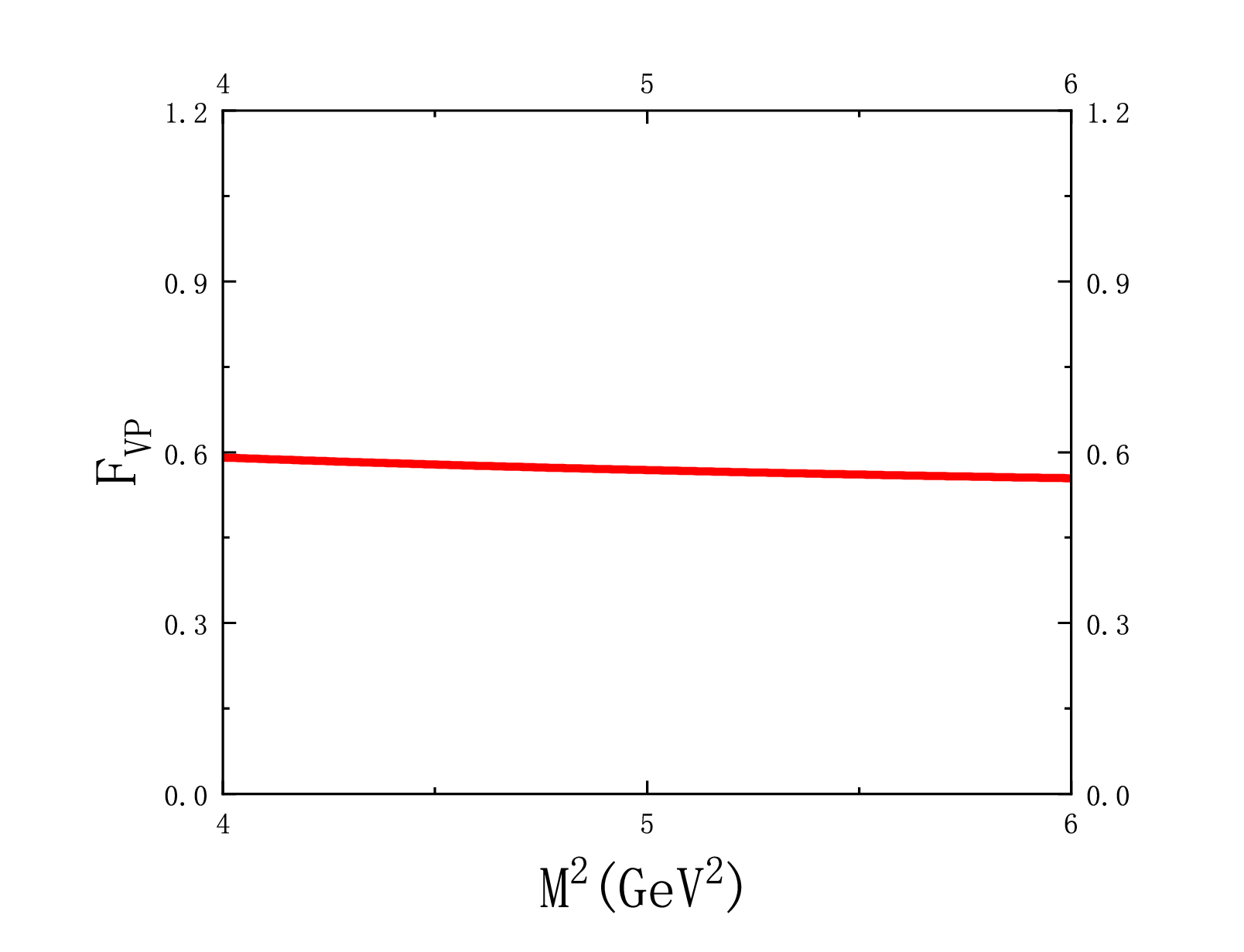}

    }
    \caption{The dependence of the form factors for $K^{*-}\rightarrow K^-\gamma$ process on the threshold $s_0$ and Borel parameter $ M^2 $.}

\end{figure}

\begin{figure}[H]
    \centering
    \vspace{-0.50cm}
    \subfigure[]{
        \includegraphics[width=0.45\linewidth]{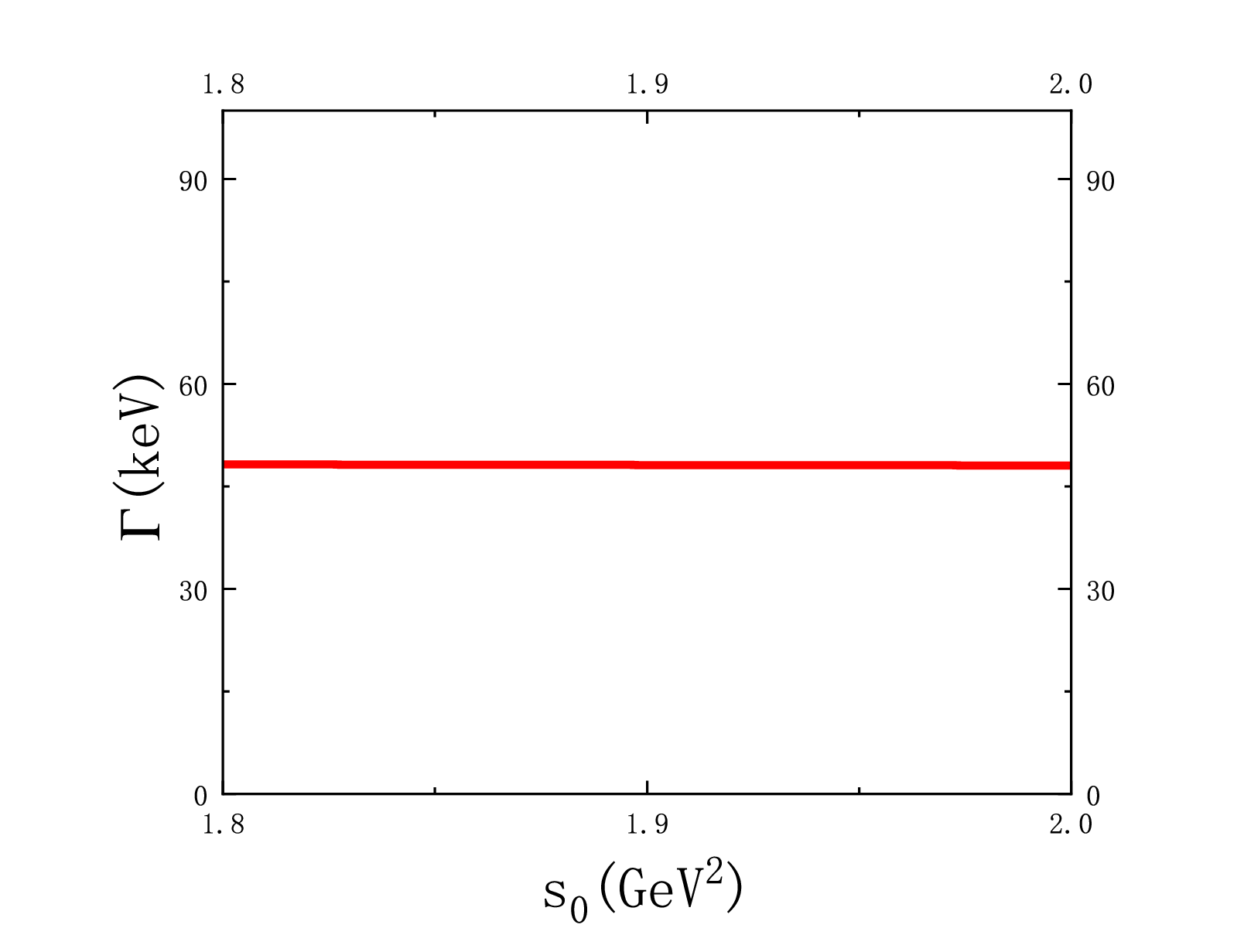}

    }
    \quad
    \subfigure[]{
        \includegraphics[width=0.45\linewidth]{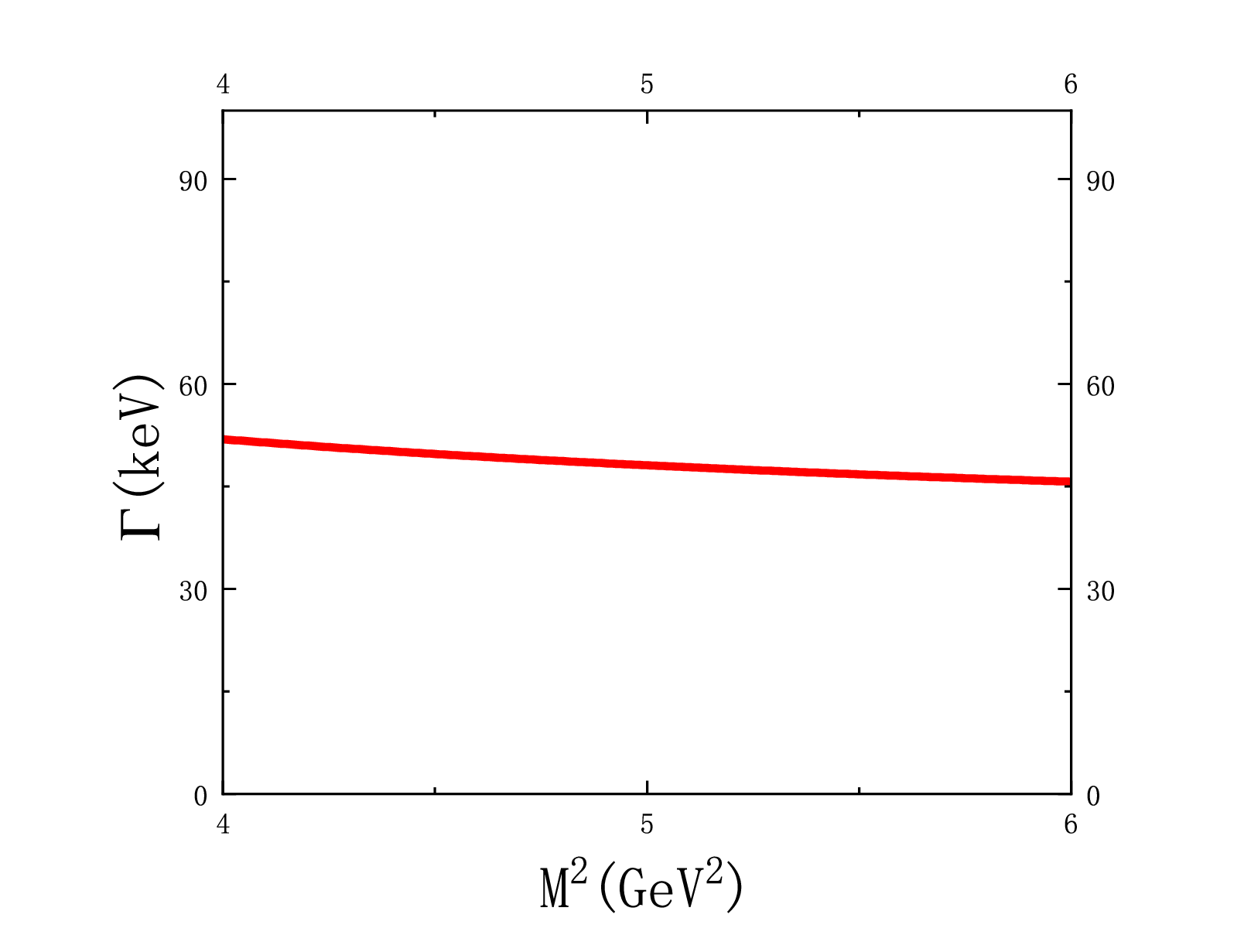}

    }
    \caption{The dependence of the decay widths for $K^{*-}\rightarrow K^-\gamma$ process on the threshold $s_0$ and Borel parameter $ M^2 $.}

\end{figure}

For the process $D^*\rightarrow D\gamma$,
\begin{figure}[H]
    \centering
    \vspace{-0.50cm}
    \subfigure[]{
        \includegraphics[width=0.45\linewidth]{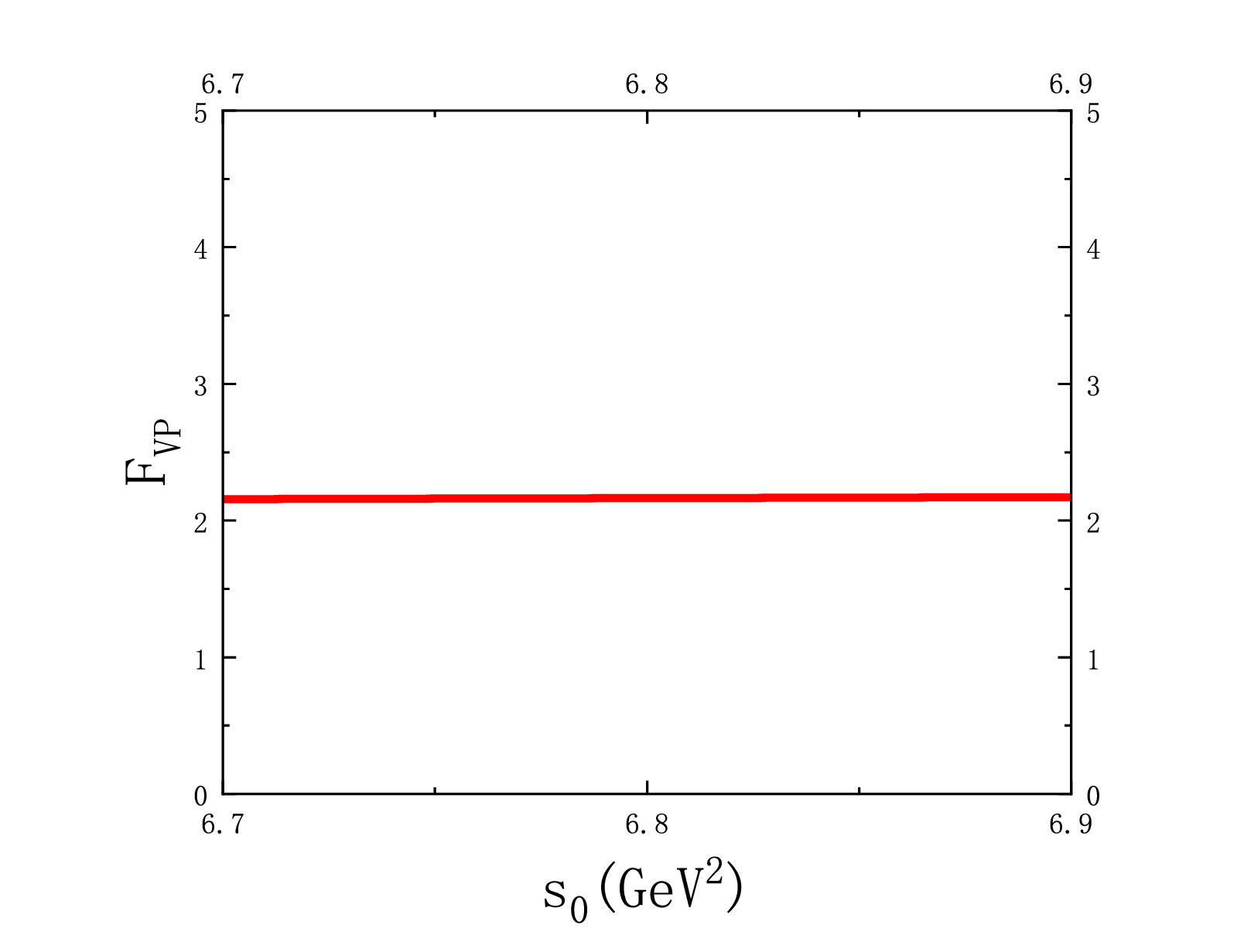}

    }
    \quad
    \subfigure[]{
        \includegraphics[width=0.45\linewidth]{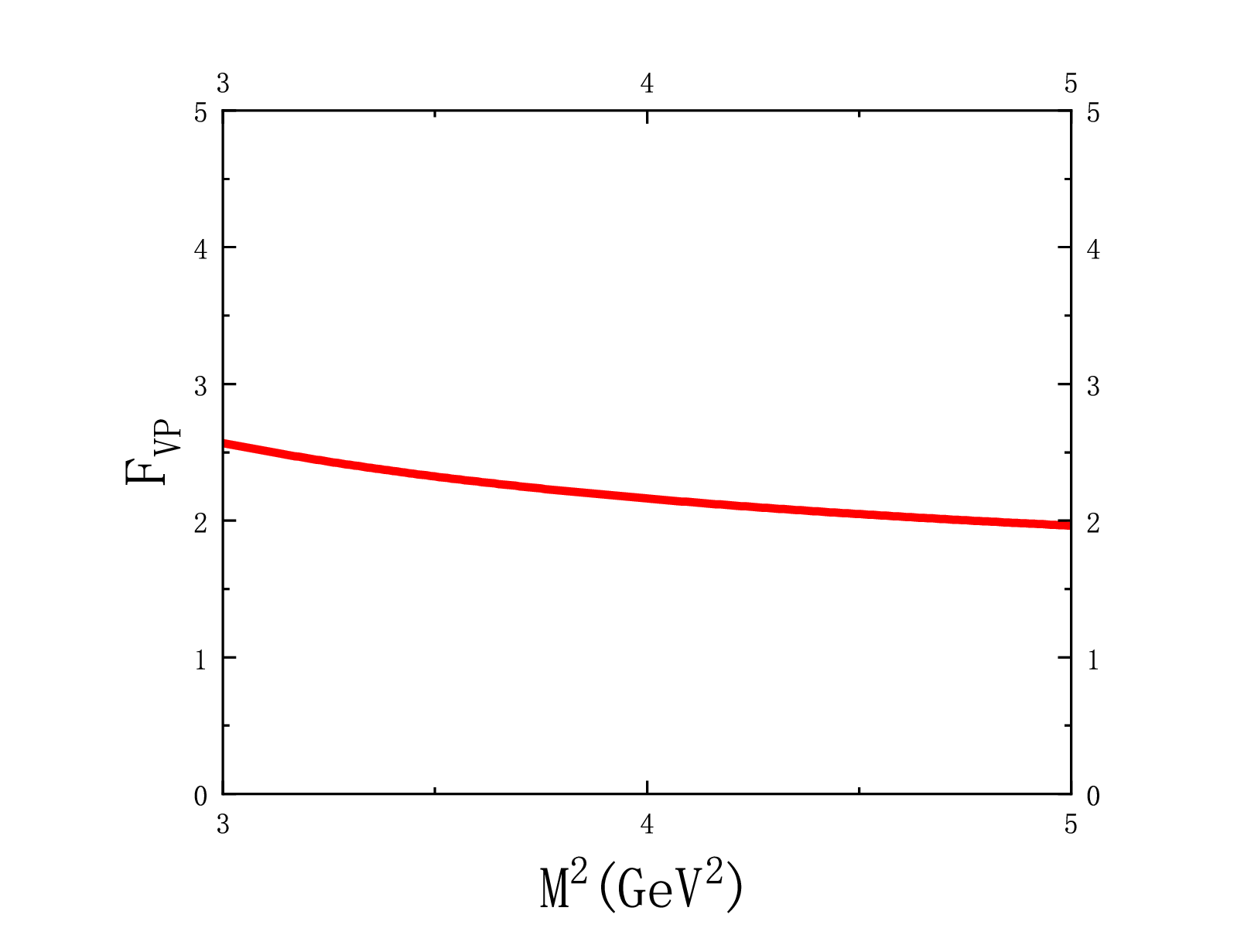}

    }
    \caption{The dependence of the form factors for $ D^*\rightarrow D\gamma $ process on the threshold $s_0$ and Borel parameter $ M^2 $.}

\end{figure}

\begin{figure}[H]
    \centering
    \vspace{-0.50cm}
    \subfigure[]{
        \includegraphics[width=0.45\linewidth]{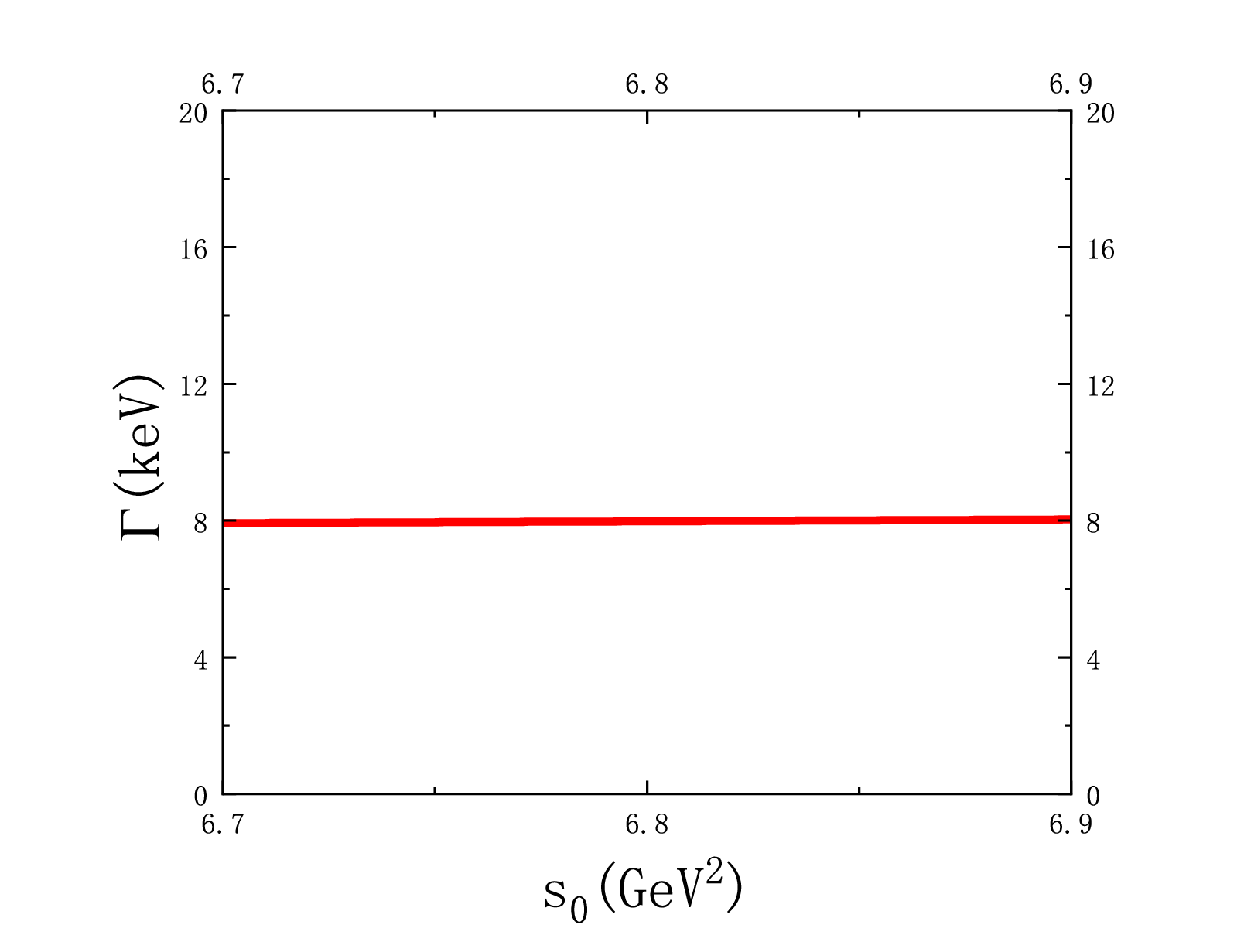}

    }
    \quad
    \subfigure[]{
        \includegraphics[width=0.45\linewidth]{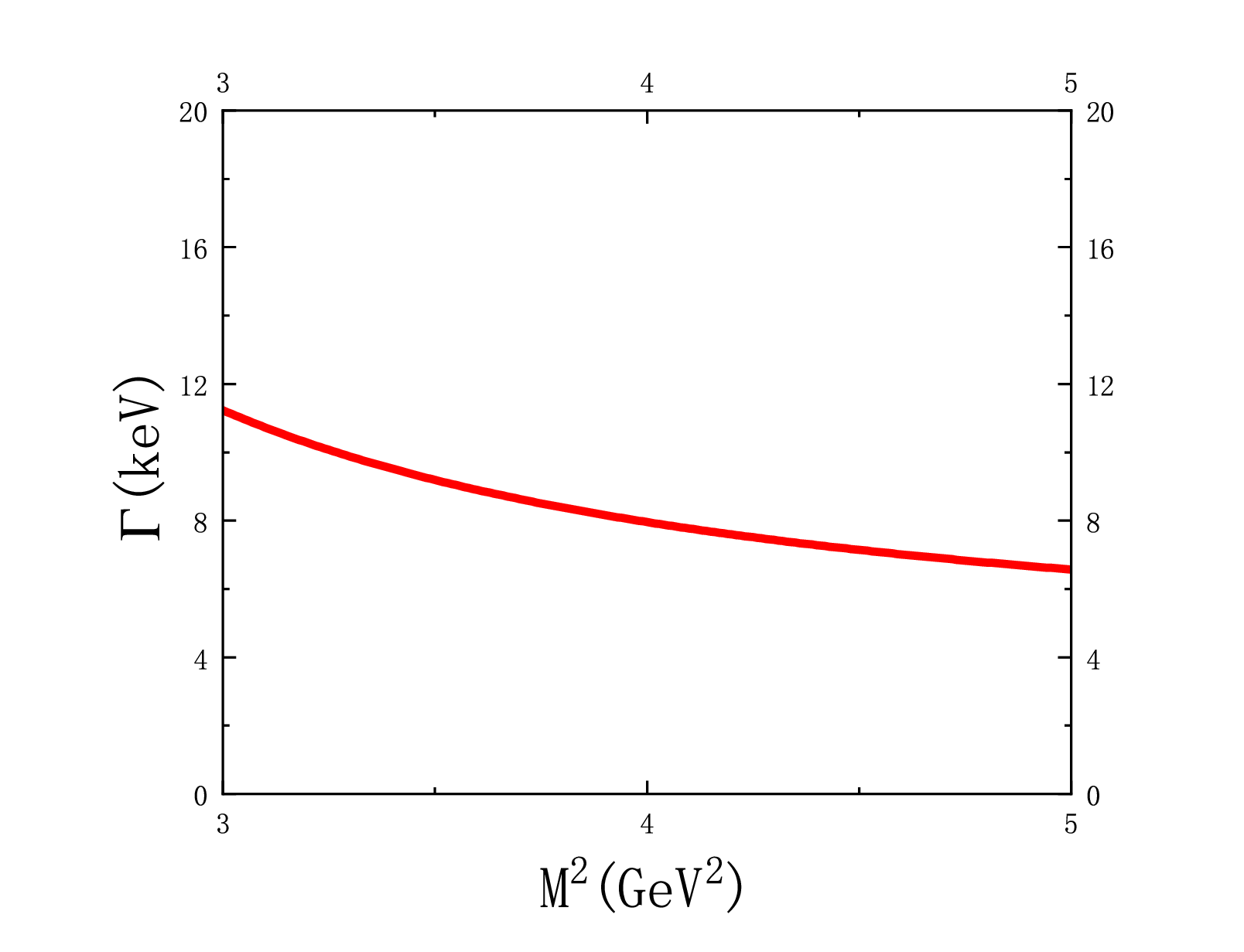}

    }
    \caption{The dependence of the decay widths for $ D^*\rightarrow D\gamma $ process on the threshold $s_0$ and Borel parameter $ M^2 $.}

\end{figure}

For the process $B^*\rightarrow B\gamma$,
\begin{figure}[H]
    \centering
    \vspace{-0.50cm}
    \subfigure[]{
        \includegraphics[width=0.45\linewidth]{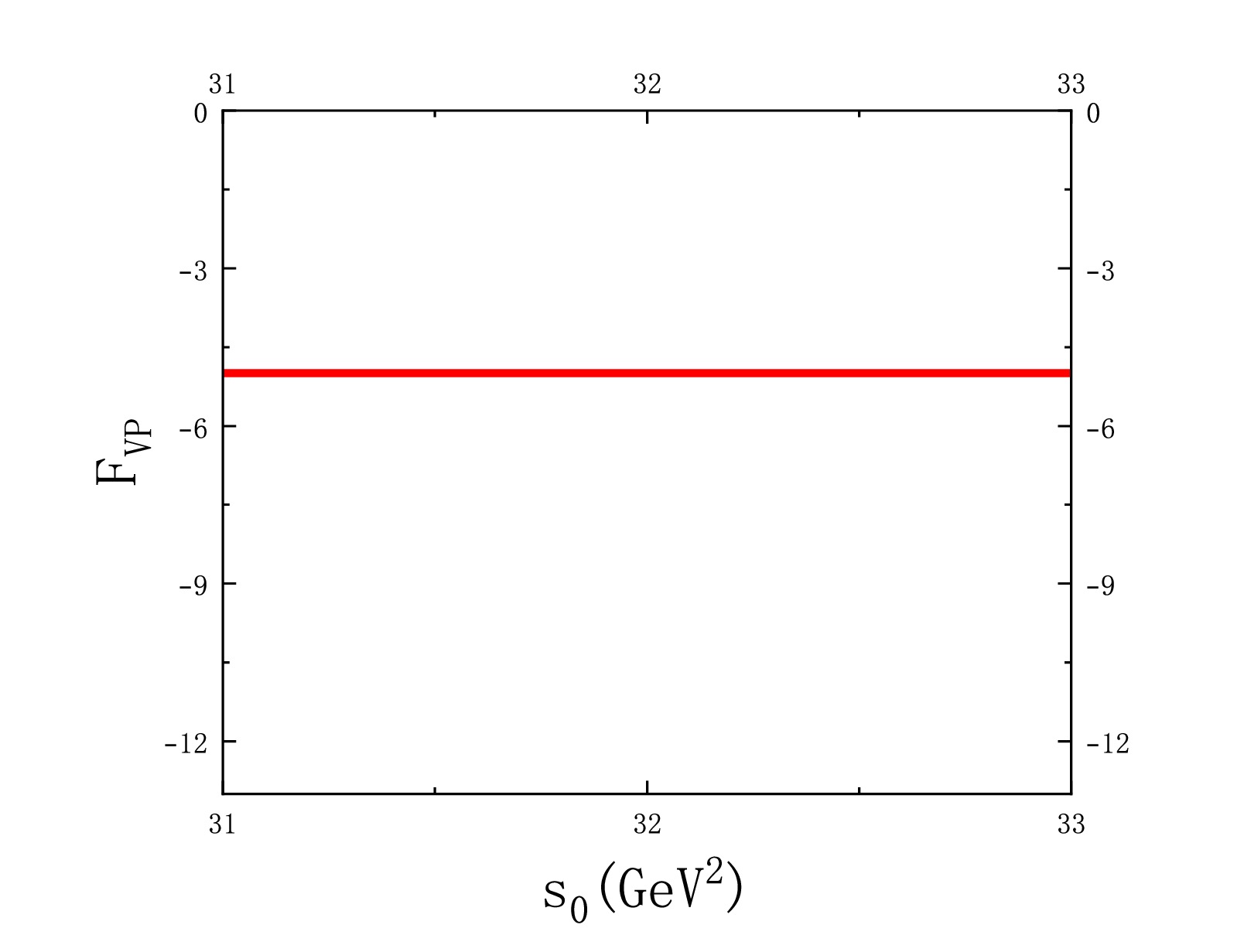}

    }
    \quad
    \subfigure[]{
        \includegraphics[width=0.45\linewidth]{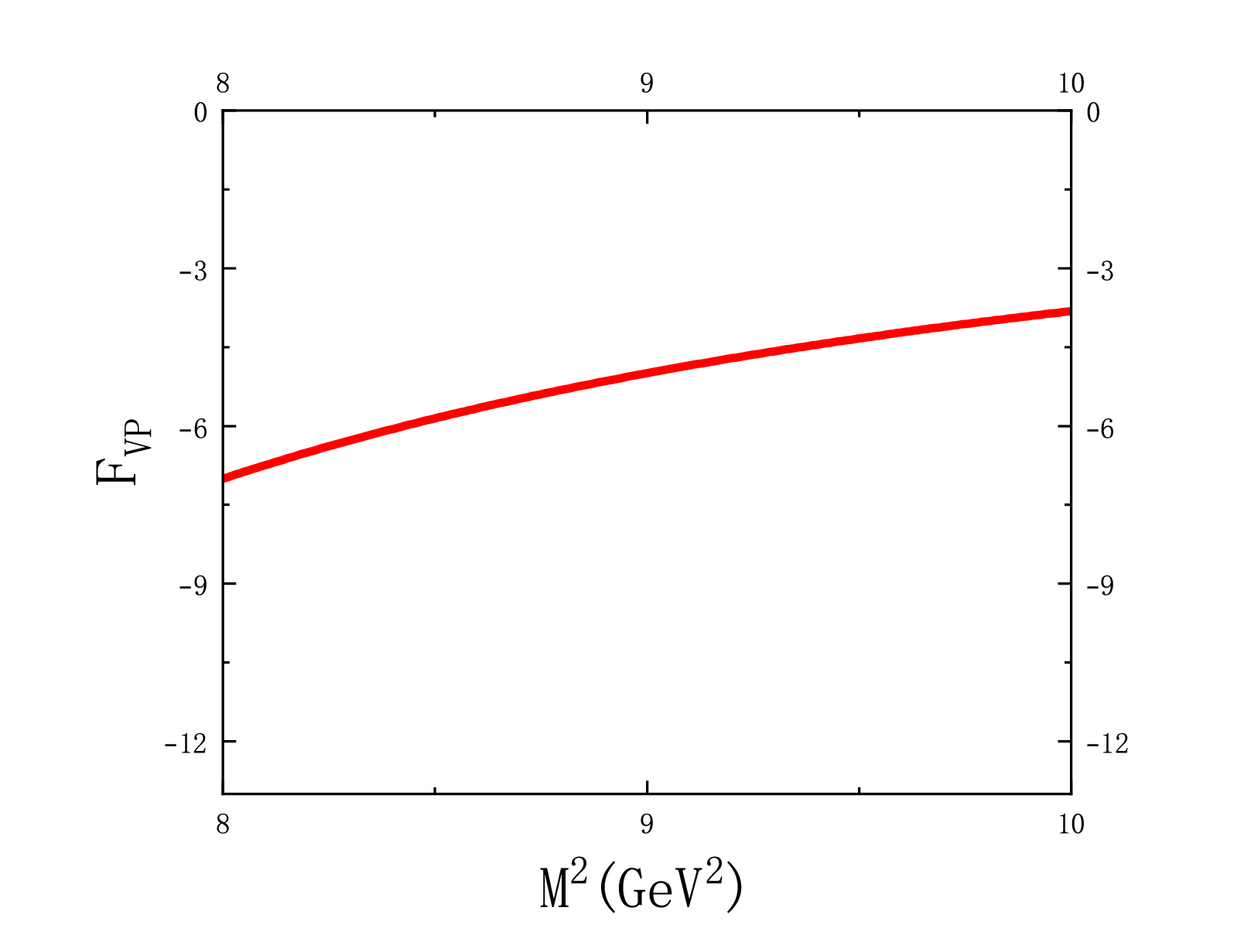}

    }
    \caption{The dependence of the form factors for $ B^*\rightarrow B\gamma $ process on the threshold $s_0$ and Borel parameter $ M^2 $.}

\end{figure}

\begin{figure}[H]
    \centering
    \vspace{-0.50cm}
    \subfigure[]{
        \includegraphics[width=0.45\linewidth]{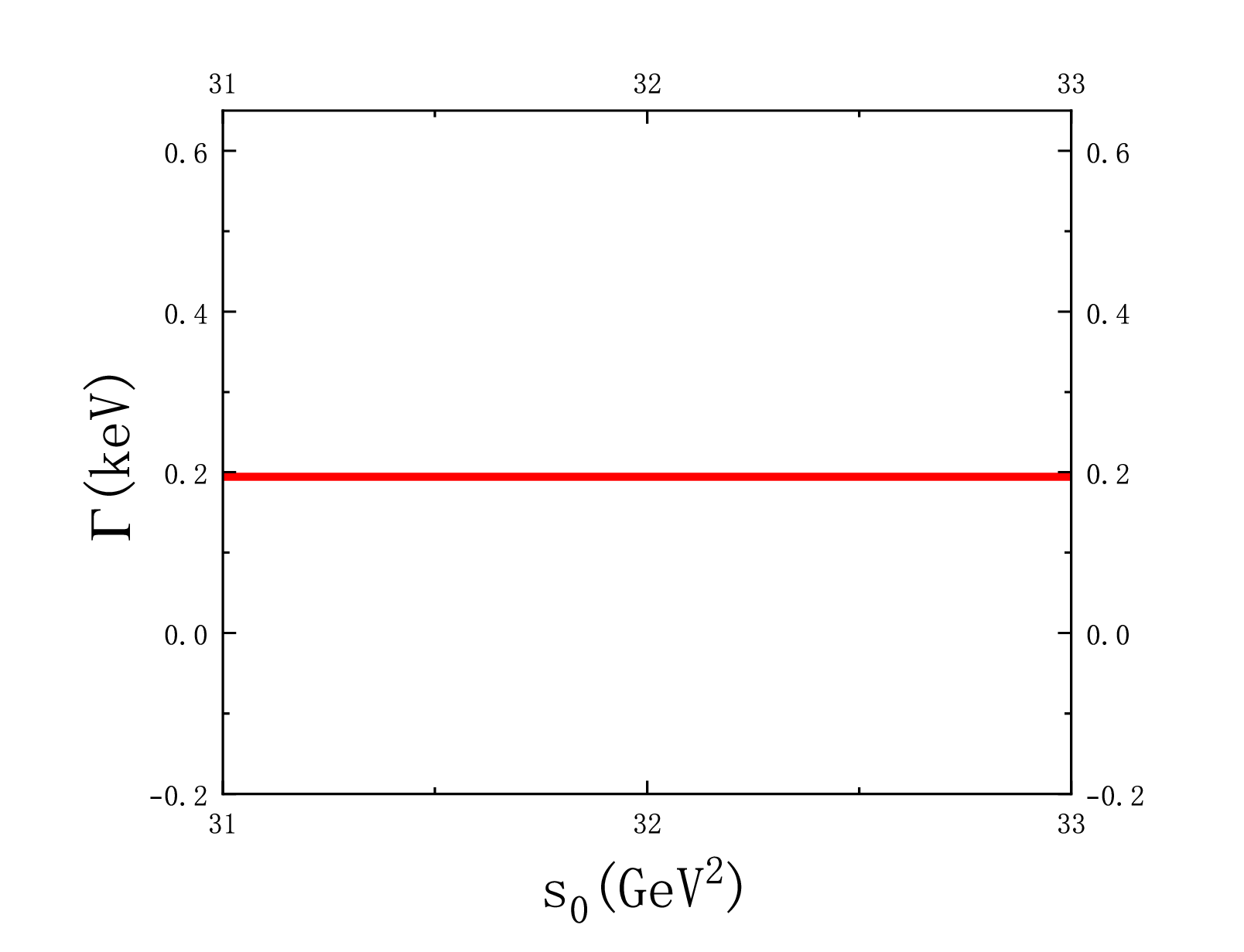}

    }
    \quad
    \subfigure[]{
        \includegraphics[width=0.45\linewidth]{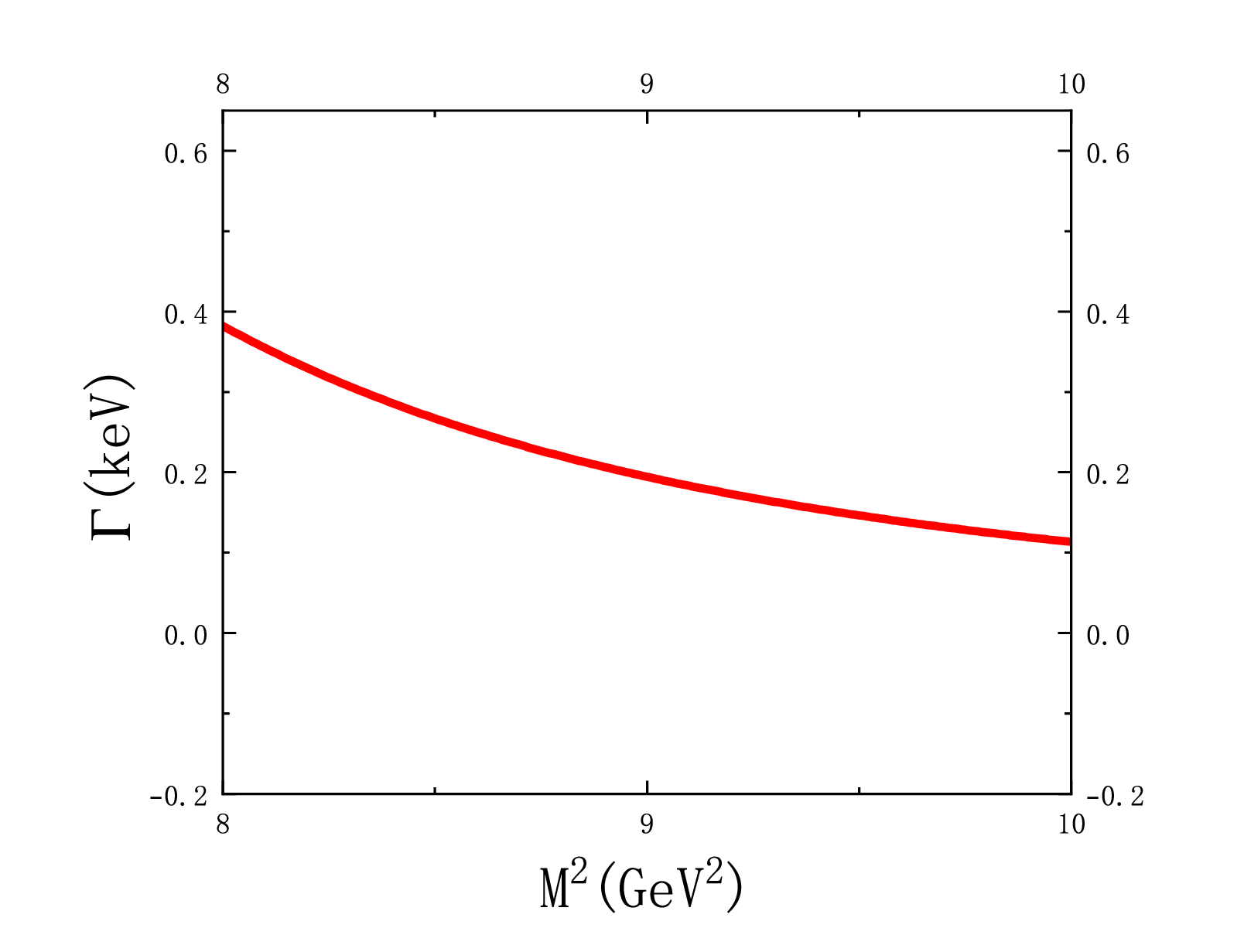}

    }
    \caption{The dependence of the decay widths for $ B^*\rightarrow B\gamma $ process on the threshold $s_0$ and Borel parameter $ M^2 $.}

\end{figure}

For the process $D^{*+}_s\rightarrow D^+_s\gamma$,
\begin{figure}[H]
    \centering
    \vspace{-0.50cm}
    \subfigure[]{
        \includegraphics[width=0.45\linewidth]{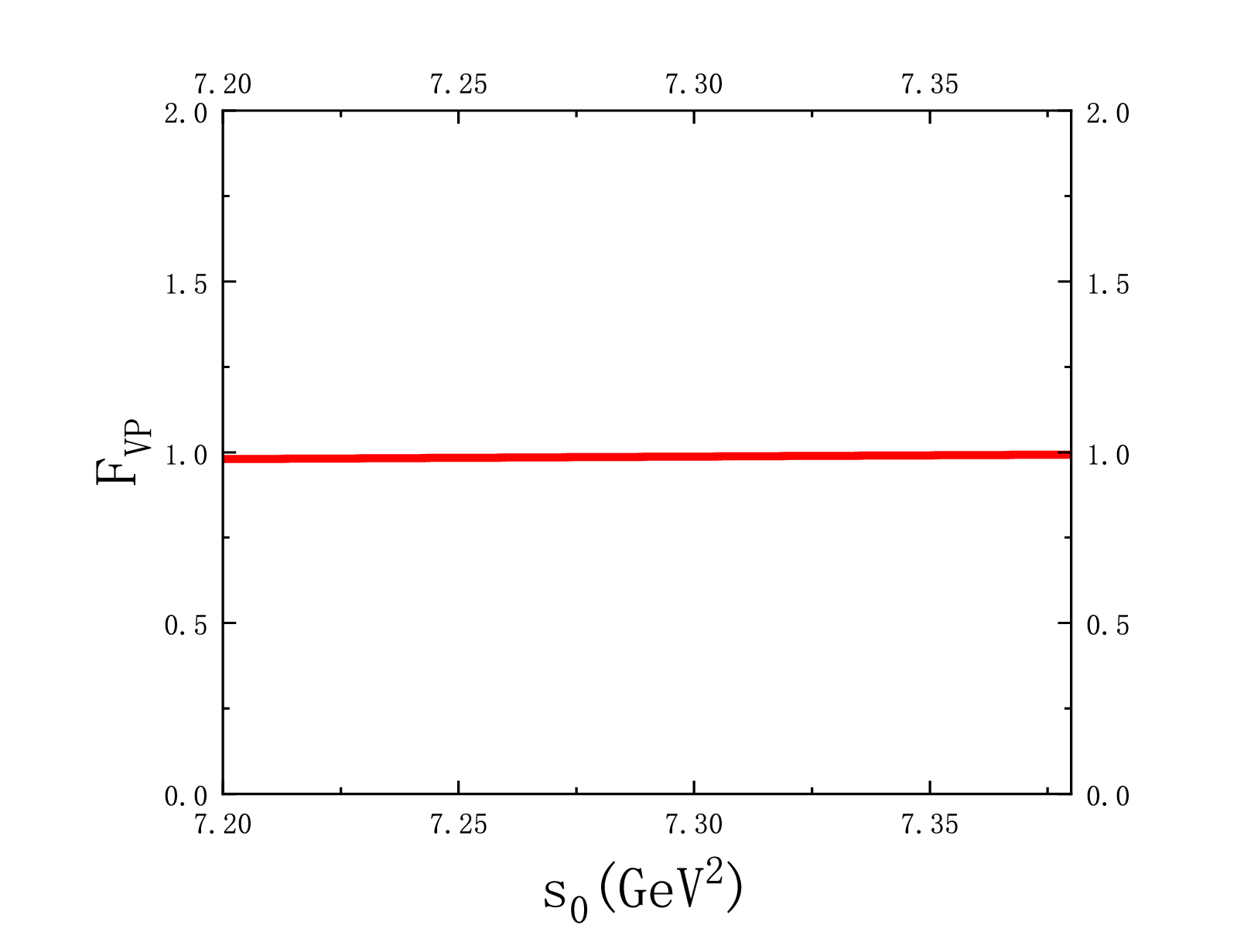}

    }
    \quad
    \subfigure[]{
        \includegraphics[width=0.45\linewidth]{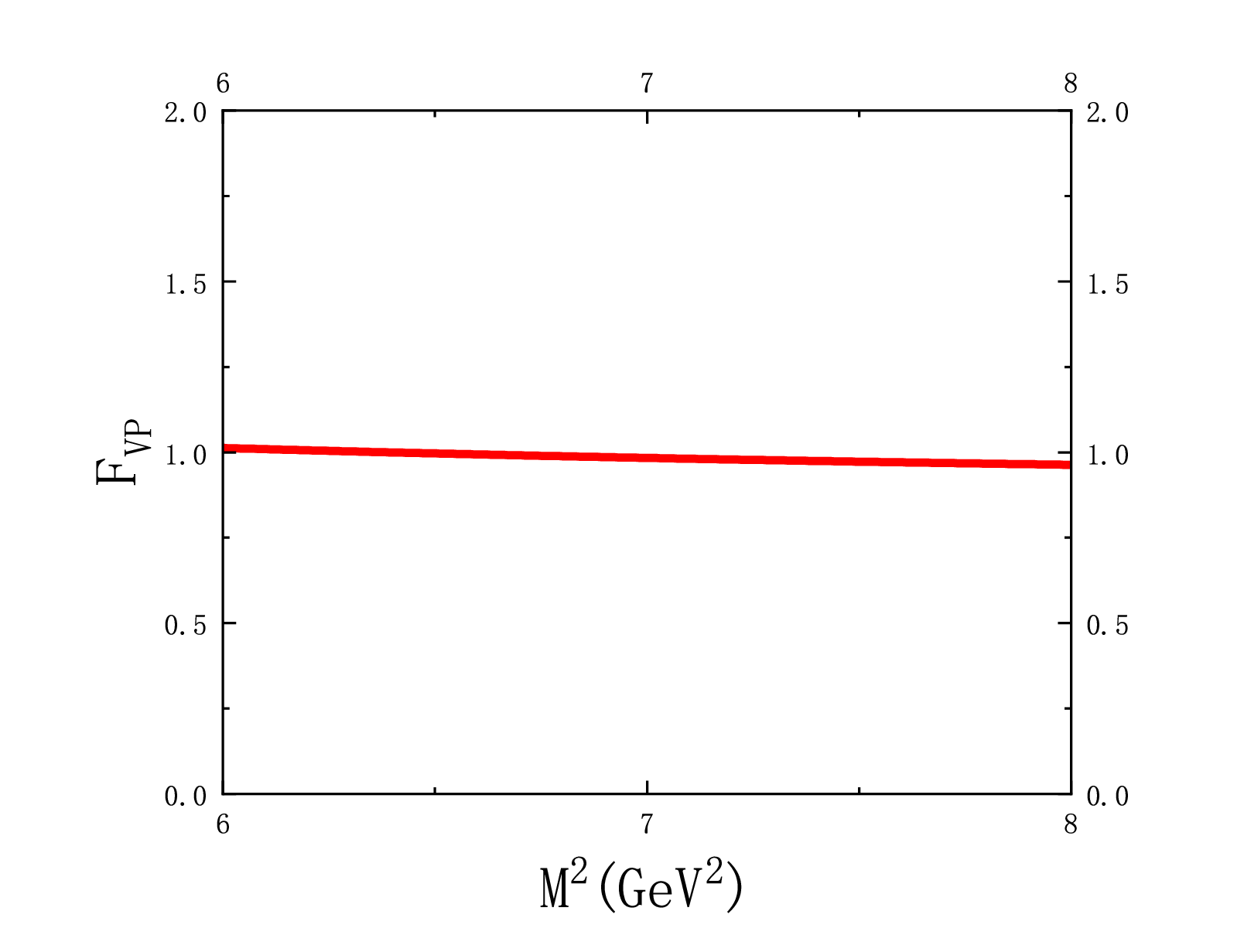}

    }
    \caption{The dependence of the form factors for $ D^{*+}_s\rightarrow D^+_s\gamma $ process on the threshold $s_0$ and Borel parameter $ M^2 $.}

\end{figure}

\begin{figure}[H]
    \centering
    \vspace{-0.50cm}
    \subfigure[]{
        \includegraphics[width=0.45\linewidth]{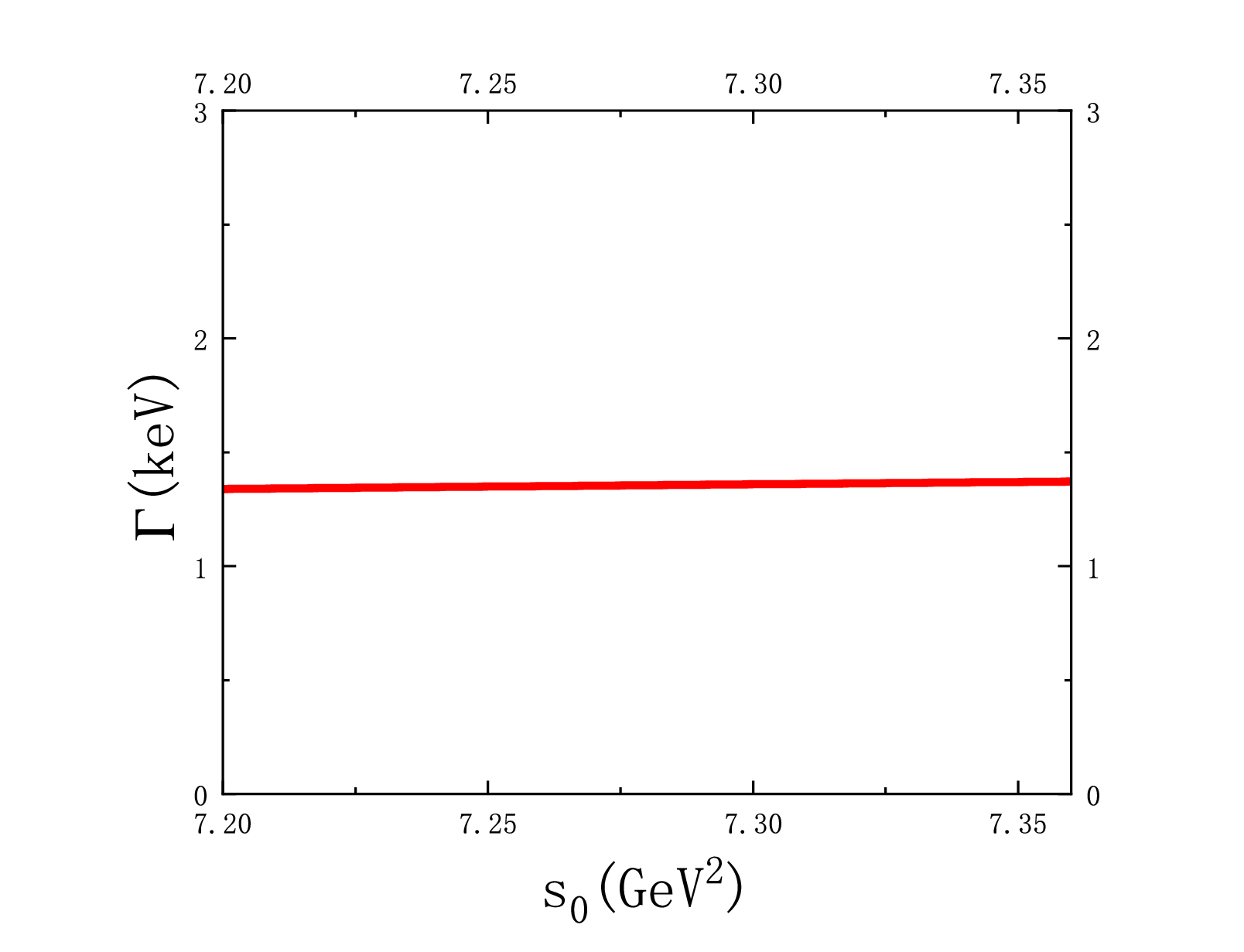}

    }
    \quad
    \subfigure[]{
        \includegraphics[width=0.45\linewidth]{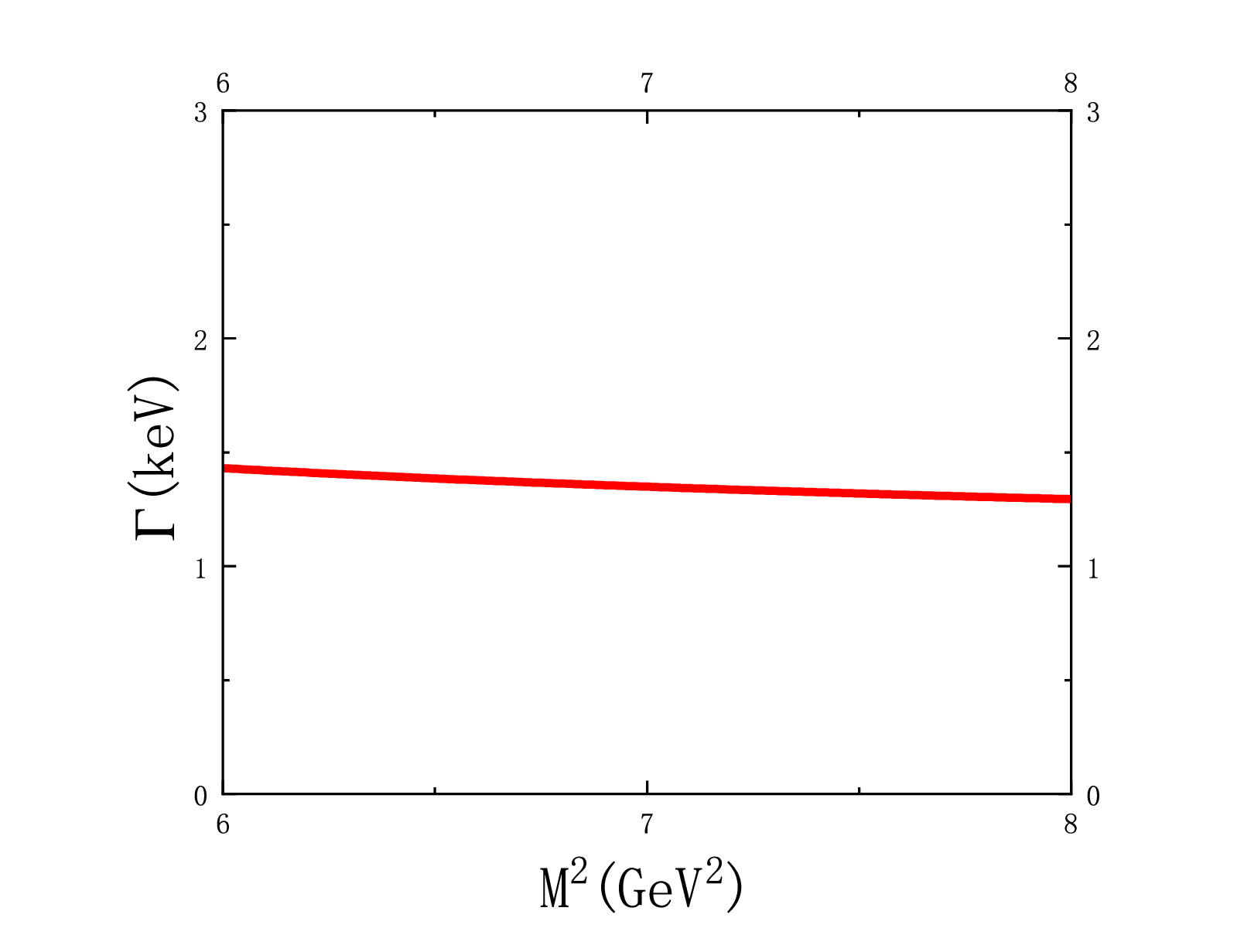}

    }
    \caption{The dependence of the decay widths for $ D^{*+}_s\rightarrow D^+_s\gamma $ process on the threshold $s_0$ and Borel parameter $ M^2 $.}

\end{figure}

For the process $B^*_s\rightarrow B_s\gamma$,
\begin{figure}[H]
    \centering
    \vspace{-0.50cm}
    \subfigure[]{
        \includegraphics[width=0.45\linewidth]{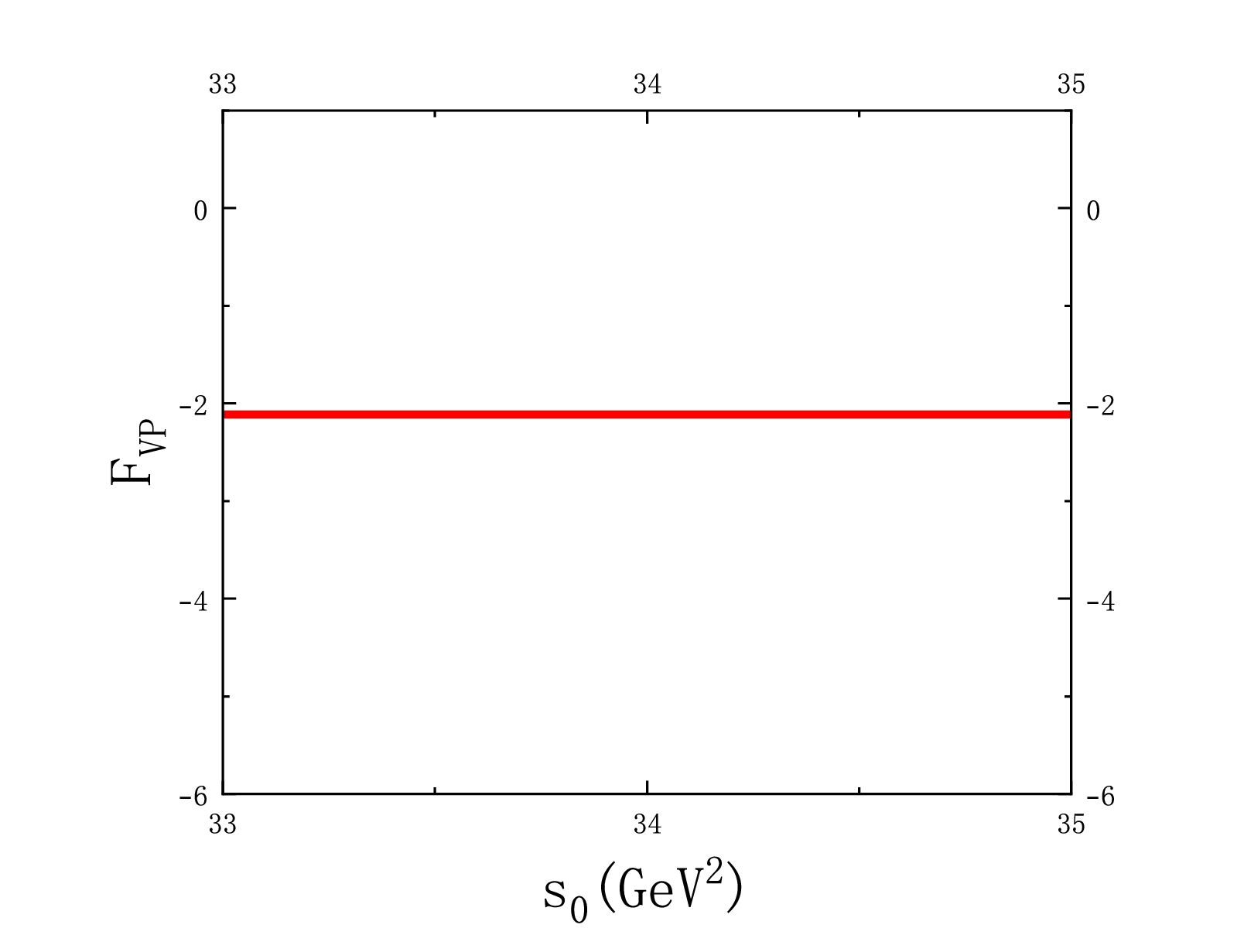}

    }
    \quad
    \subfigure[]{
        \includegraphics[width=0.45\linewidth]{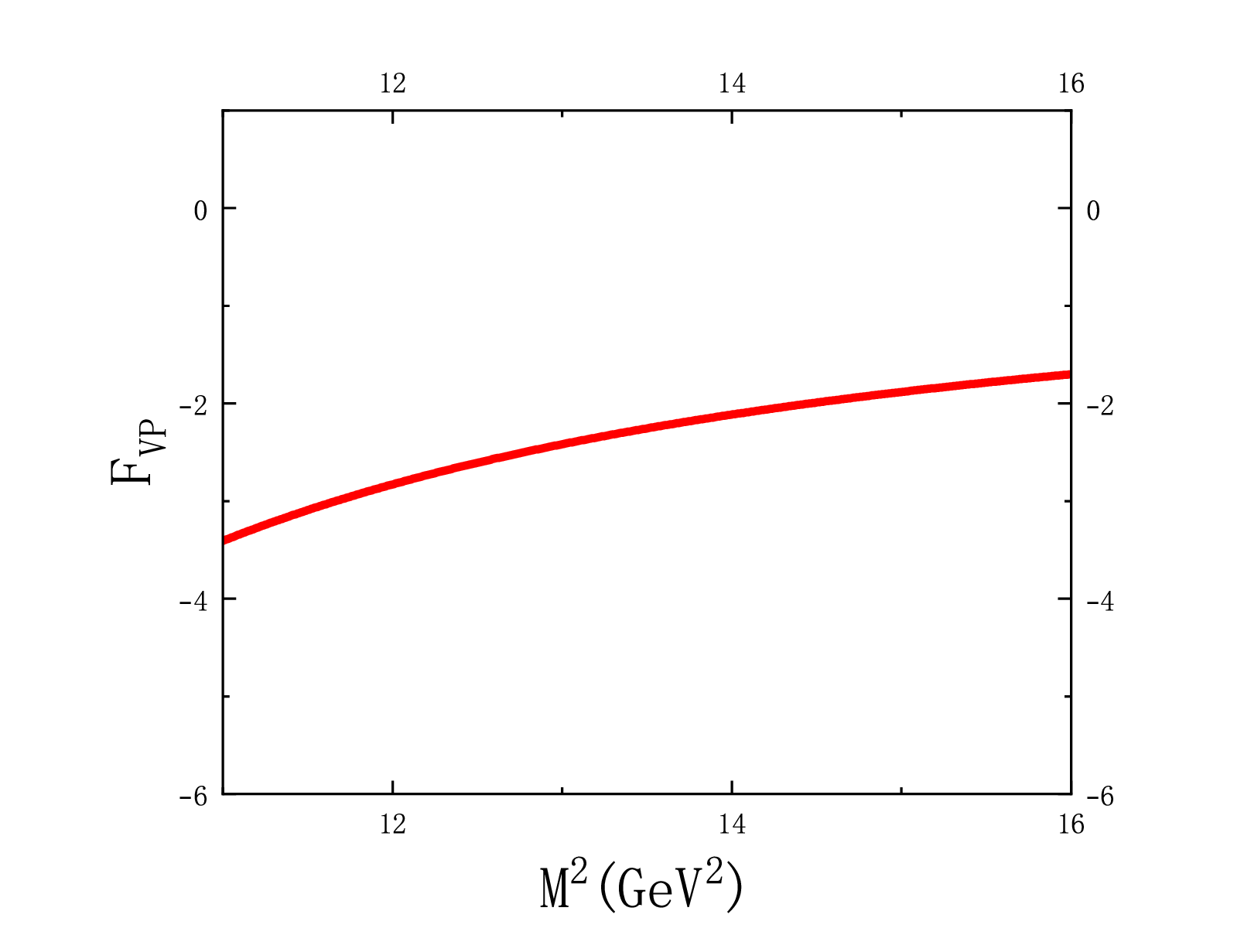}

    }
    \caption{The dependence of the form factors for $ B^*_s\rightarrow B_s\gamma $ process on the threshold $s_0$ and Borel parameter $ M^2 $.}

\end{figure}

\begin{figure}[H]
    \centering
    \vspace{-0.50cm}
    \subfigure[]{
        \includegraphics[width=0.45\linewidth]{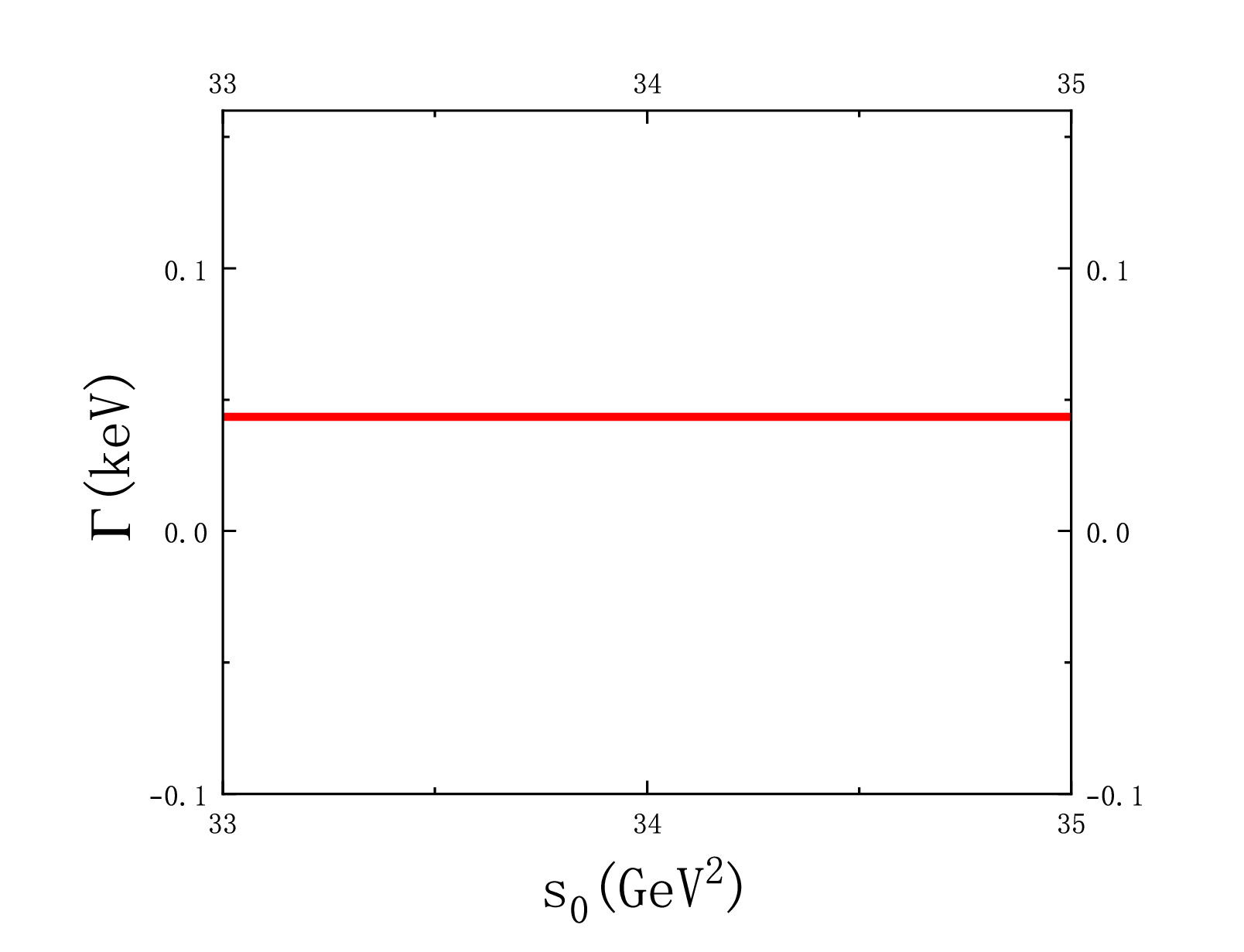}

    }
    \quad
    \subfigure[]{
        \includegraphics[width=0.45\linewidth]{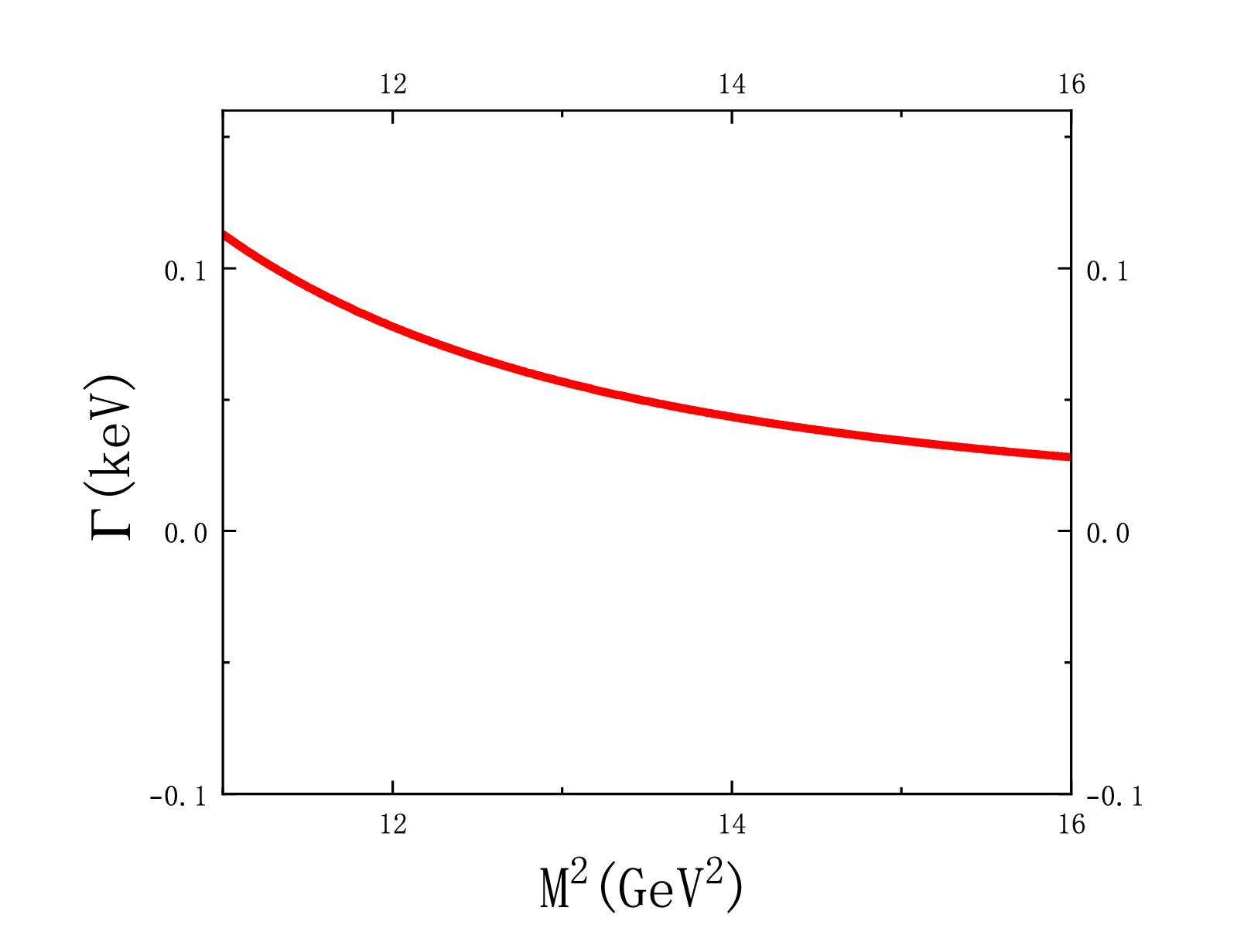}

    }
    \caption{The dependence of the decay widths for $ B^*_s\rightarrow B_s\gamma $ process on the threshold $s_0$ and Borel parameter $ M^2 $.}

\end{figure}

For the process $\psi(2S)\rightarrow\eta_c(2S)\gamma$,
\begin{figure}[H]
    \centering
    \vspace{-0.50cm}
    \subfigure[]{
        \includegraphics[width=0.45\linewidth]{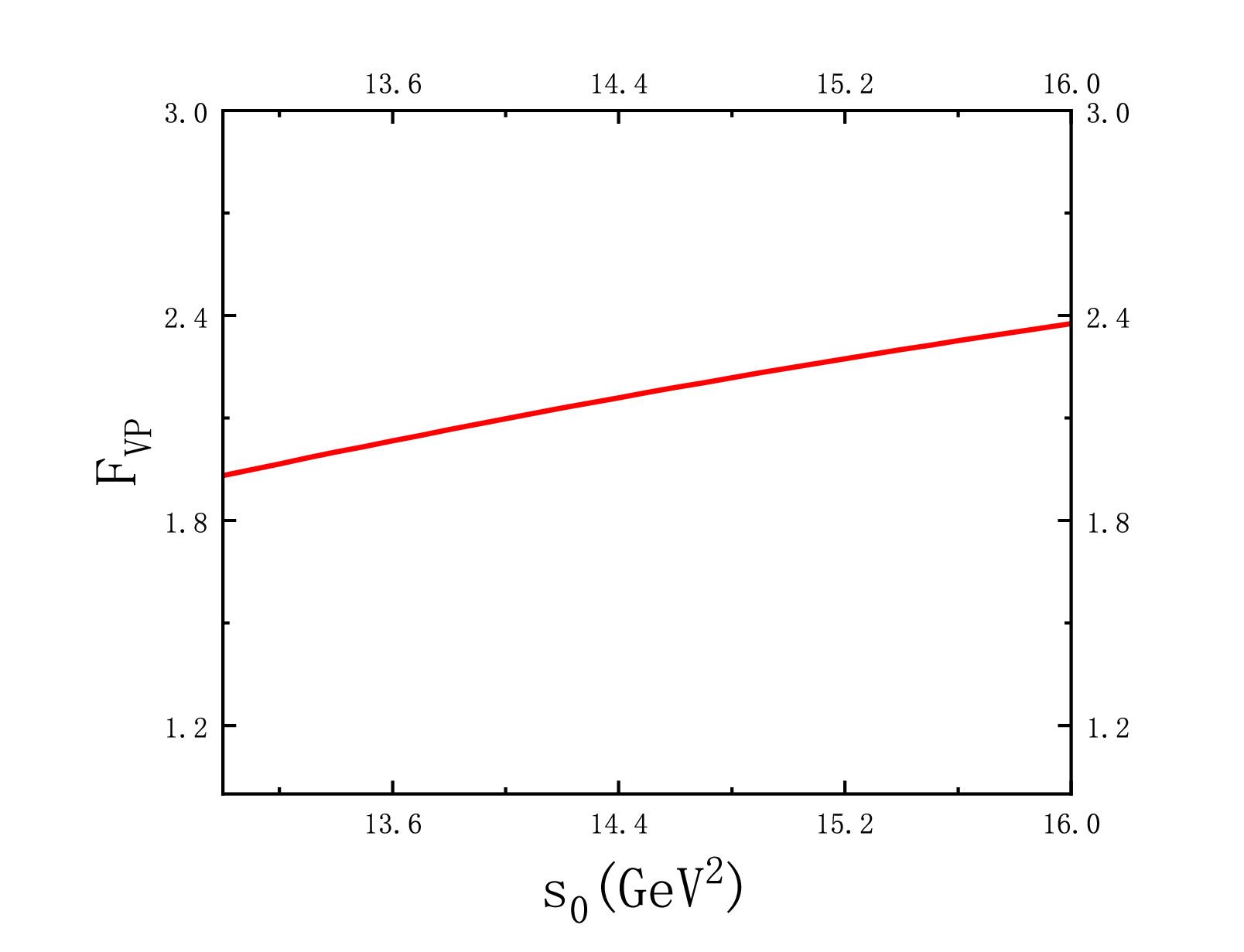}

    }
    \quad
    \subfigure[]{
        \includegraphics[width=0.45\linewidth]{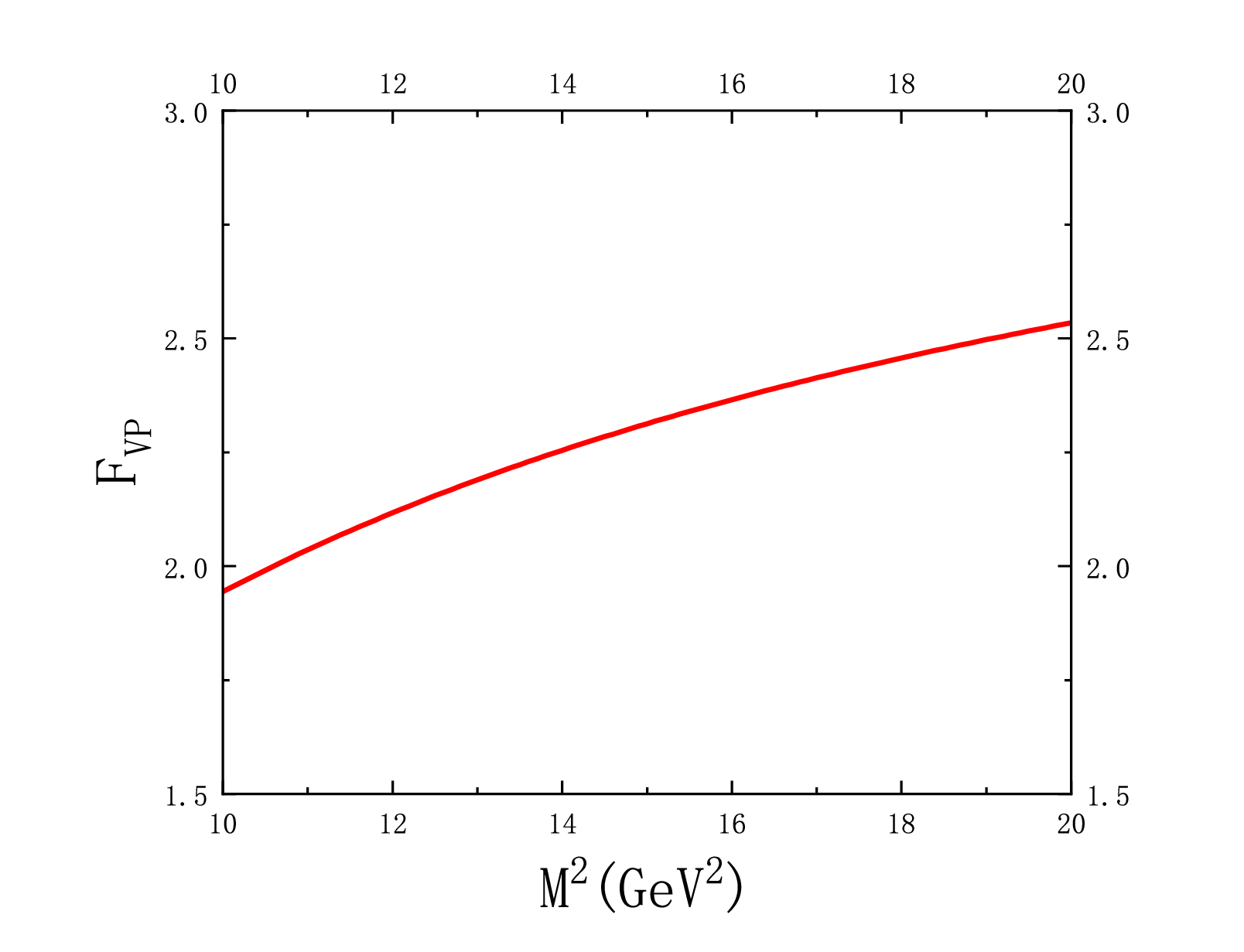}

    }
    \caption{The dependence of the form factors for $\psi(2S)\rightarrow\eta_c(2S)\gamma$ process on the threshold $s_0$ and Borel parameter $ M^2 $.}

\end{figure}

\begin{figure}[H]
    \centering
    \vspace{-0.50cm}
    \subfigure[]{
        \includegraphics[width=0.45\linewidth]{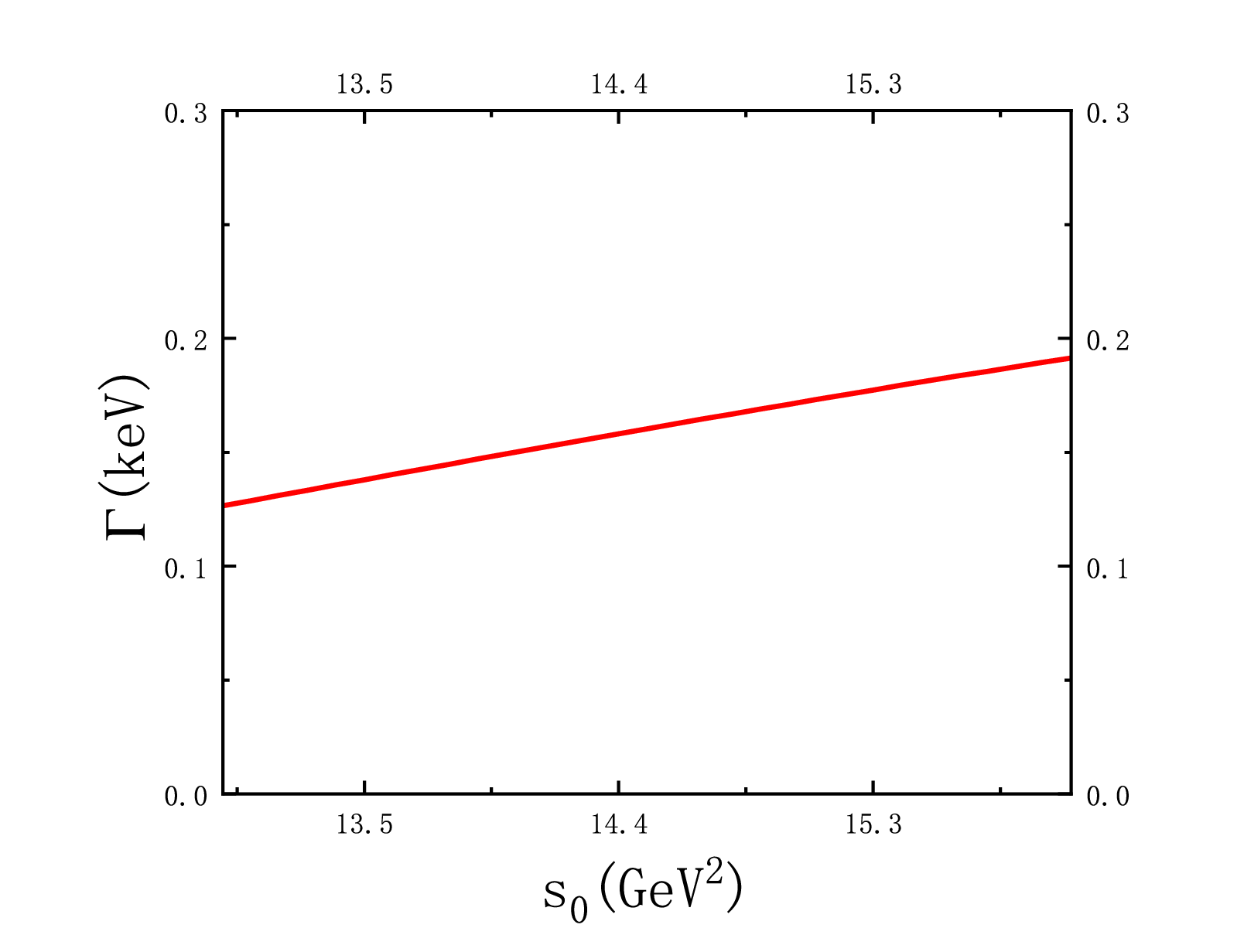}

    }
    \quad
    \subfigure[]{
        \includegraphics[width=0.45\linewidth]{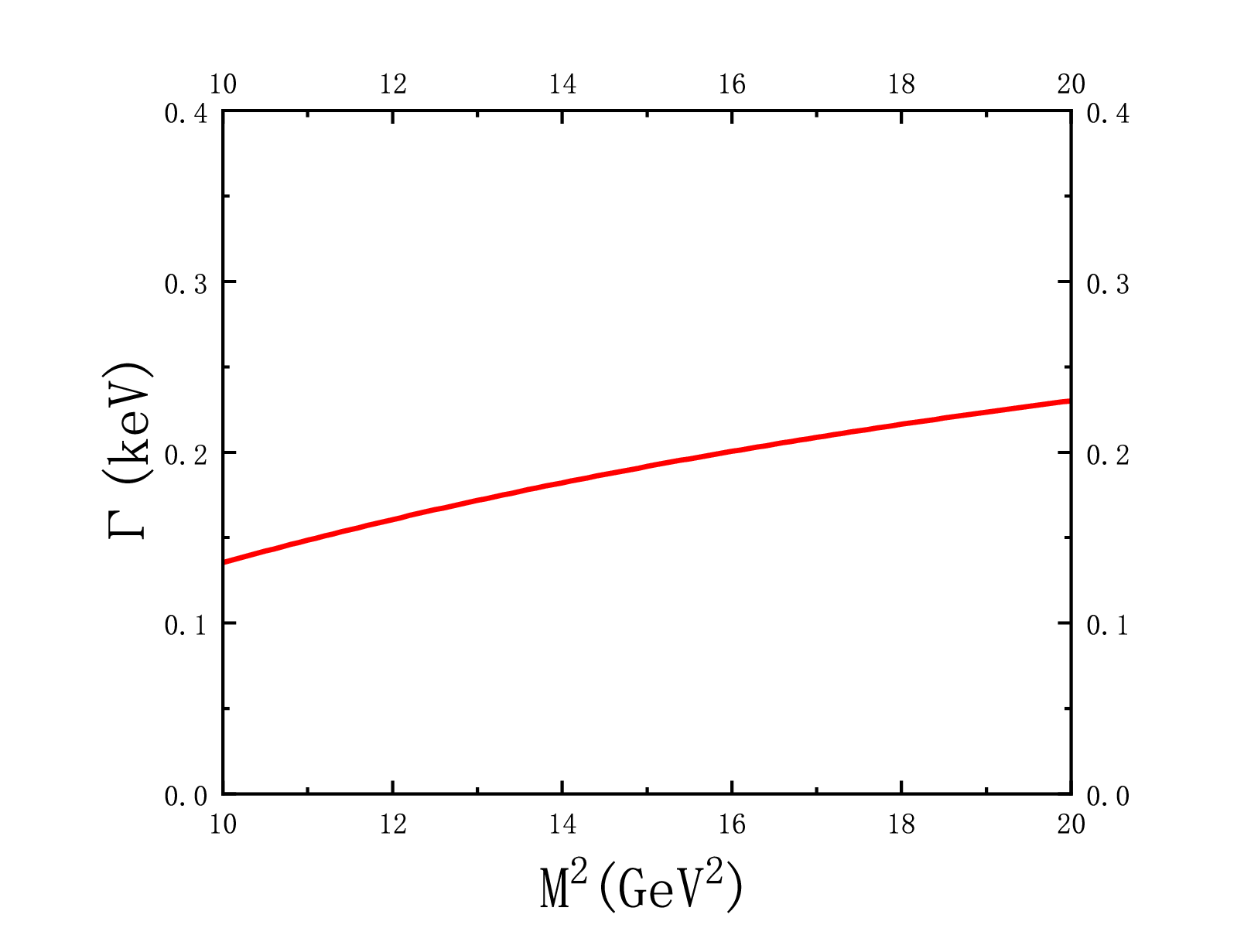}

    }
    \caption{The dependence of the decay widths for $\psi(2S)\rightarrow\eta_c(2S)\gamma$ process on the threshold $s_0$ and Borel parameter $ M^2 $.}

\end{figure}

 \bibliographystyle{apsrev4-1}
\bibliography{referen}
\label{LastBibItem}

\end{document}